\documentclass[fleqn,usenatbib]{mnras}

\usepackage{newtxtext,newtxmath}

\usepackage[T1]{fontenc}

\DeclareRobustCommand{\VAN}[3]{#2}
\let\VANthebibliography\thebibliography
\def\thebibliography{\DeclareRobustCommand{\VAN}[3]{##3}\VANthebibliography}

\usepackage{graphicx}	
\usepackage{amsmath}	
\newcommand{\TableNewLine}[2][c]{\begin{tabular}[#1]{@{}c@{}}#2\end{tabular}}
\usepackage{orcidlink}
\usepackage{comment}
\usepackage{mhchem}

\title[New Photoionization Data for \ion{Pb}{iii-vi}]{New Level Resolved Ground and Excited State Pb {\sc iii}, {\sc iv}, {\sc v} \& {\sc vi} Photoionization Cross Sections for Heavy Metal Subdwarf Modeling}

\author[D. J. Dougan et al.]{
David J. Dougan,$^{1}$\thanks{E-mail: ddougan04@qub.ac.uk}\orcidlink{0009-0008-3312-7426}
Matti Dorsch,$^2$\orcidlink{0000-0001-5400-2368}
Laura J. A. Scott,$^3$\orcidlink{0000-0001-8696-7981}
Niall E. McElroy,$^{1}$\orcidlink{0009-0002-7315-5444} \newauthor
Catherine A. Ramsbottom $^{1}$\orcidlink{0000-0003-1579-8556} and Connor P. Ballance$^{1}$\orcidlink{0000-0003-1693-1793}
\\
$^{1}$ Astrophysics Research Centre, Queen’s University Belfast, Belfast, BT7 1NN, Northern Ireland, United Kingdom\\
$^{2}$ Institut für Physik und Astronomie, Universität Potsdam, Haus 28, Karl-Liebknecht-Str. 24/25, 14476 Potsdam, Germany\\
$^3$ School of Mathematics, Statistics and Physics, Newcastle University, Newcastle upon Tyne, NE1 7RU, United Kingdom
}

\date{Accepted XXX. Received YYY; in original form ZZZ}

\pubyear{\the\year{}}

\begin{document}
\label{firstpage}
\pagerange{\pageref{firstpage}--\pageref{lastpage}}
\maketitle

\begin{abstract}
High abundances of various lead (Pb) species have been identified in the spectra of many Asymptotic Giant Branch (AGB) stars and O- and B-type subdwarfs (sdO/B).  Additional atomic data relating to Pb, and in particular photoionization cross sections, are needed to allow a greater understanding of the origin of these observed Pb abundances, and hence discern the evolutionary pathway of these stars.  We have calculated level-resolved photoionization cross sections for \ion{Pb}{iii}, {\sc iv}, {\sc v} and {\sc vi}.  Four new target structures have been developed with the General Relativistic Atomic Structure Package ({\sc grasp}$^0$), whose corresponding energy levels, Einstein A-coefficients and oscillator strengths have been found to be in good agreement with previous experimental and theoretical sources.  The photoionization cross sections calculated using the Dirac Atomic $R$-matrix Codes ({\sc darc}) are available in {\sc topbase} format, and follow the trends expected for an isonuclear series.  These new Pb data sets will now allow for the modelling of Pb abundances and line opacities under Non-Local Thermodynamic Equilibrium (non-LTE) conditions.  Using the helium-rich hot subdwarf EC\,22536$-$5304 as a test case, we show that there are noticeable differences in the Pb line profiles across the ultraviolet and optical wavelength regions under LTE and non-LTE conditions. There is both depletion and enrichment of individual Pb species.  This highlights the importance of applying non-LTE conditions when modelling EC\,22536$-$5304, as well as other O/B-type stars.

\end{abstract}




\section{Introduction}

The absorption lines of lead species (Pb, $Z$ = 82) have been observed in the ultraviolet and optical spectra of numerous stars. They are particularly strong in stars evolving into the Asymptotic Giant Branch (AGB) phase of their life.  This was first discovered in \cite{VanEck_2001}, who, using the 4057.81\AA\ \ion{Pb}{i} line, found the lead abundances to be at least 50 times greater than the solar abundance in three otherwise metal-poor stars, HD187861, HD196944 and HD224959.  Similar subsequent observations have revealed the presence of many such lead-rich giant stars \citep[e.g.][]{Johnson_2002,Sivarani_2004,Placco_2013}.  
The strong presence of the lead absorption in these stellar spectra is believed to be the result of the slow neutron-capture ($s$-) process \citep{Burbidge_1957}. During the Asymptotic Giant Branch (AGB) phase, conditions in the He-intershell create a neutron-rich environment. Protons partially mixed from the hydrogen envelope into the helium-burning layer react with $^{12}$C to form $^{13}$C via the sequence $^{12}$C($p,\gamma$)$^{13}$N($\beta^+\nu$)$^{13}$C, which then serves as the main neutron source through the $^{13}$C($\alpha$,n)$^{16}$O reaction \citep{Smith_1990, Travaglio_2004, Cui_2006}. The released neutrons are captured by seed nuclei that subsequently undergo $\beta^{-}$ decay, synthesising elements heavier than iron. This sequence continues up to the region of lead, beyond which further neutron captures produce unstable isotopes that decay back toward lead, making it a natural endpoint of the $s$-process nucleosynthesis path. 
The overall result is an accumulation of lead in the stellar composition, along with other heavy elements such as mercury \citep[Hg; e.g.,][]{Dolk_2003,Renson_2009,Gonzalez_2021,Monier_2024}, thallium \citep[Tl;][]{Leckrone_1996} and bismuth \citep[Bi;][]{Jacobs_1982,Wahlgren_2001}.  The $s$-process, along with the similar but distinct $r$-process observed in supernovae and kilonovae events \citep{Lattimer_1977,Pian_2017}, are believed to be the main processes producing elements heavier than iron in the Universe. 

High abundances of lead are also observed in hot subdwarf stars of spectral class B (sdB stars, see \cite{Heber_2016, Heber_2024} for reviews).  
These low-mass ($\approx$$0.5\,M_{\odot}$), helium-burning stars possess only a thin hydrogen-rich envelope and show chemical compositions distinct from those of normal B-type stars. Enrichment in heavy metals was first reported by \citet{O'Toole2004}, who detected strong photospheric lines of gallium, germanium, tin, and lead in several sdB stars (e.g.\ HD\,4539, HD\,171858). 
For lead in particular, the strongest absorption features in sdB spectra are typically those of \ion{Pb}{iii} and \ion{Pb}{iv}. 
The strong \ion{Pb}{iv} 1313\,\AA\ resonance line, first identified in hot subdwarf stars by \citet{O'Toole2004}, was later used to demonstrate lead enhancements of up to $\sim$1000 times the solar abundance \citep{O'Toole2006}.
A subpopulation of hot subdwarfs shows even stronger enhancements at about 10,000 times solar, exhibiting strong \ion{Pb}{iv} lines, for example at 3962.48\AA, 4049.80\AA\ and 4496.15\AA, as first shown by \cite{Naslim_2013}. 
This subset of so-called ``heavy-metal'' hot subdwarfs has evolved over time to be divided into separate sub-categories based on the dominant metal within the stellar atmospheric composition. There are several lead-rich subdwarfs \citep[e.g.][]{Naslim_2013, Jeffery2017, Wild_2017, Naslim2020, Dorsch_2021, Nemeth_2021}, as well as three zirconium-rich stars: LS\,IV$-$14$\degr$116 \citep{Naslim2011, Dorsch2020}, Feige\,46 \citep{Latour2019}, and PHL\,417 \citep{Ostensen_2020}.
Strong heavy-metal enrichment has also been observed in several hot white dwarfs \citep{Rauch2012, Rauch2020, Chayer2023}, where possible lead enrichment remains to be confirmed.

The formation of lead-rich hot subdwarfs and the origin of their enrichment remains poorly understood. Unlike heavy-metal-rich stars on the AGB and their white dwarf descendants, the enrichment observed in hot subdwarfs is unlikely to be the product of the $s$-process, since these stars are not expected to ascend the AGB. 
However, \cite{Battich_2023} and \cite{Battich2025} demonstrated that a self-synthesised route is possible when an intermediate ($i$) neutron-capture process \citep[see][]{Cowan_1977} is considered, which may take place during the first ignition of helium in these stars.  Proposed formation channels for lead- or zirconium-rich stars include the merger of two helium-core white dwarfs \citep[He-WDs, e.g.][]{Hall_2016, Schwab_2018} or the merger of a carbon-oxygen WD with a He-WD \citep{MillerBertolami_2022, Justham2011}. Two lead-rich stars have instead been found in long-period binaries with metal-poor F/G-type companions, suggesting an origin through Roche-lobe overflow at the tip of the first giant branch \citep[$P=450$ to 800\,d;][]{Dorsch_2021, Nemeth_2021}. 


As the natural end point of both the $s$- and $r$-processes, lead has proven to be a key reference for studying heavy-metal abundances in cool stars, from the Sun \citep{Helliwell_1961} and the Galactic halo \citep{Aoki_2008,Peterson_2021}, to the chemical evolution of the Galaxy \citep[e.g.][]{Contursi_2024}. 
%
Compared to these cooler stars, there are presently numerous gaps in the atomic data for many of the heavy metal and multiply charged species prominently observed in the spectra of hot subdwarf systems, including lead.   
Experimental oscillator strengths for \ion{Pb}{iii-v} have been measured \citep[e.g.][]{Andersen_1972, Ansbacher_1988, Loginov1994}, though these are few and only available for select low lying transitions.  Theoretical calculations have supplemented oscillator strengths where experimental data is not available \citep[e.g.][]{AlonsoMedina_2009,Alonso-Medina_2011,Colon_2014}.  While photoionization cross sections are available for the lower Pb charge states {\sc i} \citep[e.g.][]{Derenbach_1984,Griesmann_1991,Davidovic_2006} and {\sc ii} \citep{Muller_1990}, no such data is available for the higher charge states of Pb. 
The absence of reliable atomic data has hindered the development of accurate models for multiply ionized lead in stellar spectroscopy, thereby limiting the precision of abundance determinations. Analyses of lead-rich hot subdwarfs have so far assumed local thermodynamic equilibrium (LTE) for lead, despite its known limitations at the temperatures and densities present in these stellar photospheres \citep{Napiwotzki1997}. Consistently computed oscillator strengths and photoionization cross sections are required to enable non-LTE treatments of lead in models of hot subdwarfs, hot white dwarfs, and O/B-type stars, using model-atmosphere codes such as {\sc Tmap} \citep{Werner_2003,Werner_2012} and {\sc Tlusty/Synspec} \citep{Hubeny_2011, Lanz2003}. 
Such data would also be necessary to model atomic diffusion \citep{Michaud2011} and vertical stratification \citep{Scott_2024}  in hot subdwarf atmospheres. 

We aim to supplement the available atomic data for lead through the calculation of precise level resolved photoionization cross sections for select Pb species.  Our focus is on \ion{Pb}{iii}, {\sc iv}, {\sc v} and {\sc vi}.  These are the species which are commonly observed in the spectra of heavy metal subdwarf stars, as well as those predicted to lie in the line forming region of the stellar atmosphere.  Our intention is to improve upon the atomic data available for the stellar modeling in the atmospheres of heavy metal subdwarf stars. However, the data is intended to be multi-purpose, and is designed to be of use outside of its original purpose.  The finalized cross sections are available in {\sc topbase} format \citep{Cunto_1992} and are spectroscopically accurate to the experimental energy levels cumulated in the National Institute of Standard and Technology \citep[NIST,][]{NIST_2025} database, and to other experimental sources.

The paper is divided as follows. In Section \ref{Atomic Structures}, a discussion on the Pb target structures that act as the starting point for the subsequent collisional calculations is provided.  This includes a brief overview of the atomic structure package {\sc grasp$^0$} used to construct the target models, including our choice of configurations. For photoionization, a target structure of the residual ion is required.  Accordingly, we present four new structures for \ion{Pb}{iv}, {\sc v}, {\sc vi} and {\sc vii}.  We will discuss how well the energy levels, Einstein A-coefficients and oscillator strengths arsing from each model compare with other experimental and theoretical calculations.  In Section \ref{sec:PhotoionizationCalculations}, we provide an overview of the $R$-Matrix methodology applied to calculate the level resolved photoionization cross sections arising from our Pb models.  Sample ground and excited cross sections for each Pb species are provided. In Section \ref{sec:Application}, we show the effects of our new Pb data on the modelling of the stellar atmosphere of O- and B- type stars, using the lead-rich subdwarf star EC\,22536$-$5304 as a test case.  We illustrate the importance of applying non-LTE conditions when modelling these stars, which is now possible with the newly generated Pb photoionization data.  Section \ref{sec:Conclusions} summarises and concludes our findings.

\section{Atomic Structure - P\MakeLowercase{b} Models}
\label{Atomic Structures}

\subsection{Overview}
\label{Structure Overview}

A detailed structure of the energy levels of the residual ion is required to determine the photoionization cross sections.  Our Pb targets were developed using the General Relativistic Atomic Structure Package \citep[{\sc grasp}$^0$,][]{dyall_1989}, which employs a Dirac-Coulomb Hamiltonian to solve the time independent Dirac equation.  {\sc grasp}$^0$ has been extensively described in other publications and will not be described in detail here.  For further background, please refer to publications such as \cite{Dougan_2025} and \cite{McCann_2025}.

We present the four finalised structure models \ion{Pb}{iv} (Section \ref{sec:Structure-PbIV}), {\sc v} (Section \ref{sec:Structure-PbV}), {\sc vi} (Section \ref{sec:Structure-PbVI}) and {\sc vii} (Section \ref{sec:Structure-PbVII}).  In addition, we will discuss how the energies, Einstein A-coefficients and oscillator strengths arsing from our structures compare with experimental and theoretical equivalents presented in NIST and other sources.

\subsection{\ion{Pb}{iv} Structure}
\label{sec:Structure-PbIV}

Three times ionized lead consists of 79 electrons, making it a part of the gold isoelectronic sequence.  It has a ground state of [Xe]4f$^{14}$5d$^{10}$($^2$S$_{1/2}$) and an ionization potential of 3.111 Ryd \citep{Hanni_2010}.   Our {\sc grasp}$^{0}$ model for \ion{Pb}{iv} consists of 19 orbitals going up to $n$=7 and $l$=3.  The orbitals include 1s, 2s, 2p, 3s, 3p, 3d, 4s, 4p, 4d, 4f, 5s, 5p, 5d, 6s, 6p, 6d, 7s, 7p and 7d.  There are 22 unique configurations included in the target, with the critical ones being the single electron promotion out of the 6s and 5d$^{10}$ orbitals.  Promotions out of the 5s$^2$ and 5p$^6$ orbitals are included to better align the target energies to those in the literature. Table \ref{tab:Pb_IV_Structure} lists the finalised configuration set used.  In total, there are 691 energy levels in our \ion{Pb}{iv} target. 

   \begin{table}
   
     \centering
        \caption[]{The configurations included in the wavefunction expansion for the {\sc grasp$^{0}$} structure of \ion{Pb}{IV}.}
         \begin{tabular}{p{2cm} p{2cm } p{2cm}}
            \hline
            \noalign{\smallskip}
            \multicolumn{3}{l}{\textbf{\ion{Pb}{iv} - 22 Configurations}} \\
            \noalign{\smallskip}
            \hline
            \noalign{\smallskip}
              5s$^{2}$5p$^{6}$5d$^{10}$6s       & 5s$^{2}$5p$^{6}$5d$^{10}$6p       & 5s$^{2}$5p$^{6}$5d$^{10}$6d             \\ \noalign{\smallskip}
              5s$^{2}$5p$^{6}$5d$^{10}$7s       & 5s$^{2}$5p$^{6}$5d$^{10}$7p       & 5s$^{2}$5p$^{6}$5d$^{10}$7d             \\ \noalign{\smallskip}
              5s$^{2}$5p$^{6}$5d$^{9}$6s$^{2}$  & 5s$^{2}$5p$^{6}$5d$^{9}$6s6p      & 5s$^{2}$5p$^{6}$5d$^{10}$6s6d           \\ \noalign{\smallskip}
              5s$^{2}$5p$^{6}$5d$^{9}$6s7s      & 5s$^{2}$5p$^{6}$5d$^{9}$6s7p      & 5s$^{2}$5p$^{6}$5d$^{9}$6s7d            \\ \noalign{\smallskip}
              5s$^{2}$5p$^{6}$5d$^{9}$6p$^{2}$  & 5s$^{2}$5p$^{6}$5d$^{9}$6d$^{2}$  & 5s5p$^{6}$5d$^{9}$6s6d$^{2}$            \\ \noalign{\smallskip}
              5s$^{2}$5p$^{5}$5d$^{10}$6s6p     & 5s$^{2}$5p$^{5}$5d$^{10}$6s7p     & 5s$^{2}$5p$^{4}$5d$^{10}$6s6p$^{2}$     \\ \noalign{\smallskip}
              5s5p$^{6}$5d$^{10}$6s7p           & 5s5p$^{6}$5d$^{10}$6d7d           & 5s5p$^{6}$5d$^{10}$6d$^{2}$           \\ \noalign{\smallskip}
              5p$^{6}$5d$^{10}$6s7d$^{2}$       & {}                                & {}                                      \\ \noalign{\smallskip}
            \hline
            \noalign{\smallskip}
         \end{tabular}
        
        \label{tab:Pb_IV_Structure}
   \end{table}

The NIST database contains 108 experimental energy levels for \ion{Pb}{iv}, obtained from the works of \cite{Moore_1971}, \cite{Gutmann_1973} \& \cite{Raassen_1991}.  We compared the lowest lying 59 levels that have an equivalent in our \ion{Pb}{iv} target, the first 20 of which are highlighted in Table \ref{tab:PbIV-Energy-Levels}.  There is very good agreement in the energy levels between our target structure and those observed experimentally.  The differences tend to be $\leq$ 0.40 Ryd.  Levels 2, 3 and 4 show very high percentage deviation compared to the rest of the levels, but this is to be expected given the smaller magnitudes involved.  Their absolute energy deviations are similar to the other energy levels.  The average relative percentage difference across all 59 levels we compared was found to be $-0.741$\%.

\begin{table}
  \caption{A comparison of the first 20 energy levels in the \ion{Pb}{iv} target.  The average relative percentage difference between all the shifted energy levels is $-0.741$\%. The NIST energies for the first 20 energy levels were obtained from \protect\cite{Moore_1971} and \protect\cite{Gutmann_1973}.}
  \centering
\resizebox{0.99\columnwidth}{!}{
    \begin{tabular}{ccccc} \hline
            \textbf{Level} & \textbf{Config.}      & \TableNewLine{\textbf{NIST} \\ \textbf{/ Ryd}} & \TableNewLine{\textbf{GRASP$^0$} \\ \textbf{/ Ryd}} & \TableNewLine{\textbf{Percentage} \\ \textbf{Error / \%}} \\ \hline
 \noalign{\smallskip}
    1     &  5d$^{10}$6s ($^2$S$_{1/2}$)                & 0.00000 & 0.00000 & \\ \noalign{\smallskip}
    2     &  5d$^{10}$6p ($^2$P$_{1/2}^{\circ}$)        & 0.69400 & 0.86288 & 24.335 \\ \noalign{\smallskip}
    3     &  5d$^{10}$6p ($^2$P$_{3/2}^{\circ}$)        & 0.88592 & 1.02337 & 15.515 \\ \noalign{\smallskip}
    4     &  5d$^{9}$6s$^{2}$ ($^2$D$_{5/2}$)           & 0.92267 & 0.96951 & \phantom{-}5.076 \\ \noalign{\smallskip}
    5     &  5d$^{9}$6s$^{2}$ ($^2$D$_{3/2}$)           & 1.11692 & 1.15173 & \phantom{-}3.117 \\ \noalign{\smallskip}
    6     &  5d$^{9}$6s6p ($^4$P$_{5/2}^{\circ}$)       & 1.51606 & 1.48900 & \phantom{-}1.785 \\ \noalign{\smallskip}
    7     &  5d$^{9}$6s6p ($^4$F$_{7/2}^{\circ}$)       & 1.57346 & 1.55880 & \phantom{-}0.931 \\ \noalign{\smallskip}
    8     &  5d$^{9}$6s6p ($^4$F$_{5/2}^{\circ}$)       & 1.57876 & 1.56214 & \phantom{-}1.052 \\ \noalign{\smallskip}
    9     &  5d$^{9}$6s6p ($^4$P$_{3/2}^{\circ}$)       & 1.59826 & 1.58632 & \phantom{-}0.747 \\ \noalign{\smallskip}
    10    &  5d$^{10}$6d  ($^2$D$_{3/2}$)               & 1.68183 & 1.64398 & \phantom{-}2.251 \\ \noalign{\smallskip}
    11    &  5d$^{10}$7s  ($^2$S$_{1/2}$)               & 1.68679 & 1.72143 & \phantom{-}2.054 \\ \noalign{\smallskip}
    12    &  5d$^{10}$6d  ($^2$D$_{5/2}$)               & 1.70240 & 1.66254 & \phantom{-}2.342 \\ \noalign{\smallskip}
    13    &  5d$^{9}$6s6p  ($^4$F$_{3/2}^{\circ}$)      & 1.72010 & 1.68344 & \phantom{-}2.131 \\ \noalign{\smallskip}
    14    &  5d$^{9}$6s6p  ($^4$F$_{9/2}^{\circ}$)      & 1.72981 & 1.68379 & \phantom{-}2.661 \\ \noalign{\smallskip}
    15    &  5d$^{9}$6s6p  ($^2$D$_{3/2}^{\circ}$)      & 1.76319 & 1.74482 & \phantom{-}1.042 \\ \noalign{\smallskip}
    16    &  5d$^{9}$6s6p  ($^2$F$_{5/2}^{\circ}$)      & 1.76582 & 1.74221 & \phantom{-}1.337 \\ \noalign{\smallskip}
    17    &  5d$^{9}$6s6p  ($^4$D$_{7/2}^{\circ}$)      & 1.76654 & 1.75120 & \phantom{-}0.868 \\ \noalign{\smallskip}
    18    &  5d$^{9}$6s6p  ($^4$P$_{1/2}^{\circ}$)      & 1.76921 & 1.73517 & \phantom{-}1.924 \\ \noalign{\smallskip}
    19    &  5d$^{9}$6s6p  ($^4$F$_{5/2}^{\circ}$)      & 1.79542 & 1.78049 & \phantom{-}0.831 \\ \noalign{\smallskip}
    20    &  5d$^{9}$6s6p  ($^2$P$_{3/2}^{\circ}$)      & 1.82273 & 1.80600 & \phantom{-}0.918 \\ \noalign{\smallskip}
    \hline
    \noalign{\smallskip}
    \end{tabular}%
}
    \label{tab:PbIV-Energy-Levels}%
\end{table}

\begin{figure}
\centering
\includegraphics[width=0.9\columnwidth]{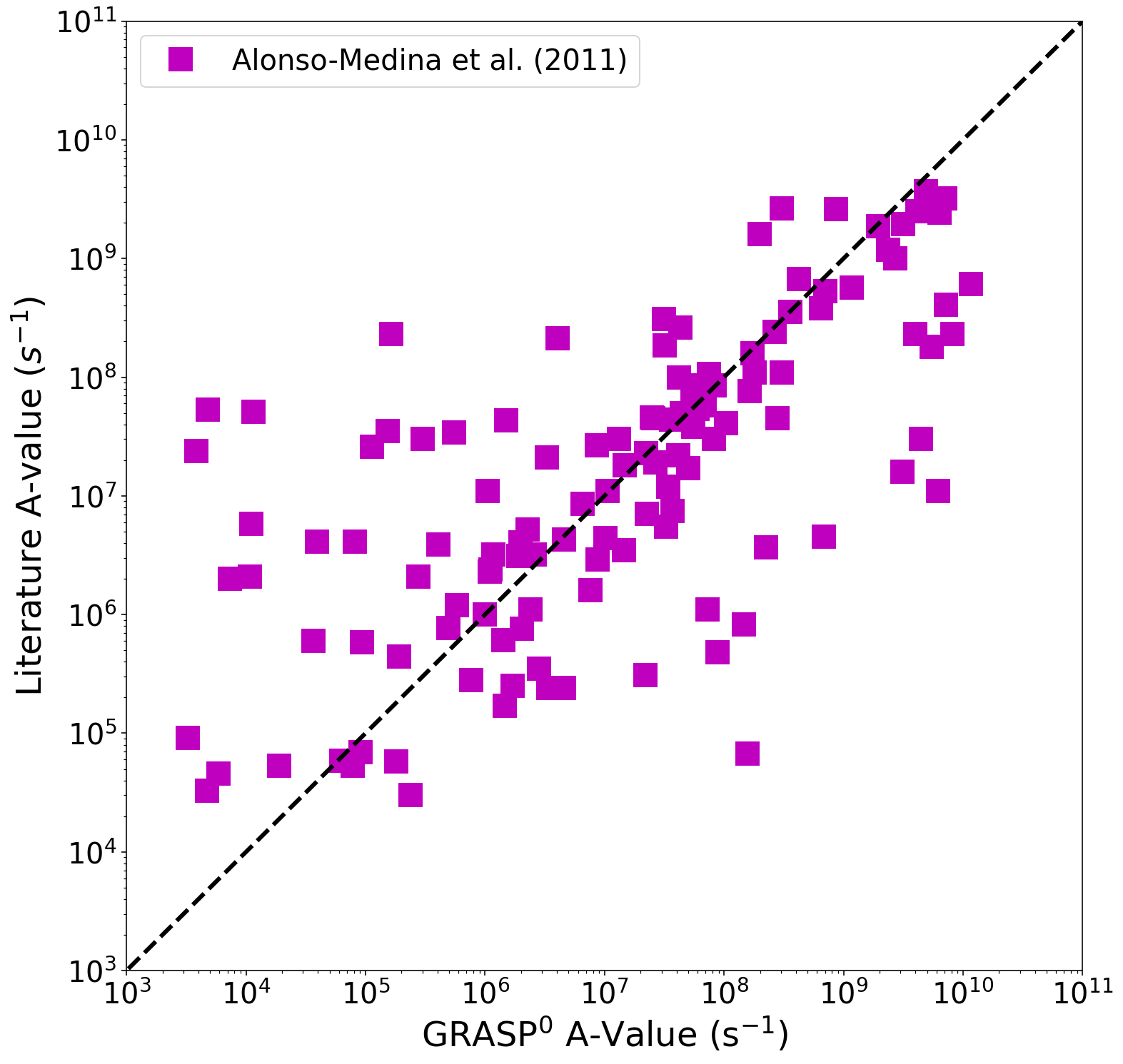}
\caption{A comparison between the A-values arising from our \ion{Pb}{iv} model and those computed in \protect\cite{Alonso-Medina_2011}.}
\label{fig:PbIV_A-values}
\end{figure}

For the A-values, we compare our results with those of \citet{Alonso-Medina_2011}, who employed the methods of \citet{Cowan_1981}. Specifically, we examined 124 of their transitions and compared them to our calculations, where both the upper and lower energy levels of each transition were adjusted to the spectroscopic values listed in NIST.
See Fig.\ \ref{fig:PbIV_A-values} for a graphical illustration of this comparison.  The majority of the compared A-values are in good agreement with our {\sc grasp$^0$} run and those computed in \cite{Alonso-Medina_2011}, though there are a few which exhibit discrepancies by several orders of magnitude.  The difference in calculation methods could explain these discrepancies.  The Cowan code applies a semi-relativistic Hartree-Fock approach with core polarisation effects, in contrast to our fully relativistic {\sc grasp$^0$} calculations.   

An additional test we can perform to confirm the validity of the data is to make a comparison of the oscillator strengths ($f$).  This fractional number is a measure of the number of electrons available for a transition.  For an electric dipole (E1) transition between lower level $i$ and upper level $k$, the corresponding oscillator strength is defined as:

\begin{equation}
    f_{ik} = \frac{2 (\Delta E)}{3g_i}S
\end{equation}

where $(\Delta E)$ is the energy separation between the lower and upper levels in the transition, $g_i$ is the statistical weight of the lower level, and $S$ is the line strength determined from the corresponding transition matrix element.  As for the A-values, the oscillator strength is sensitive to the energy separation and as such the oscillator strengths are recalculated using spectroscopically accurate experimentally determined energy levels.

\begin{figure}
\centering
\includegraphics[width=0.9\columnwidth]{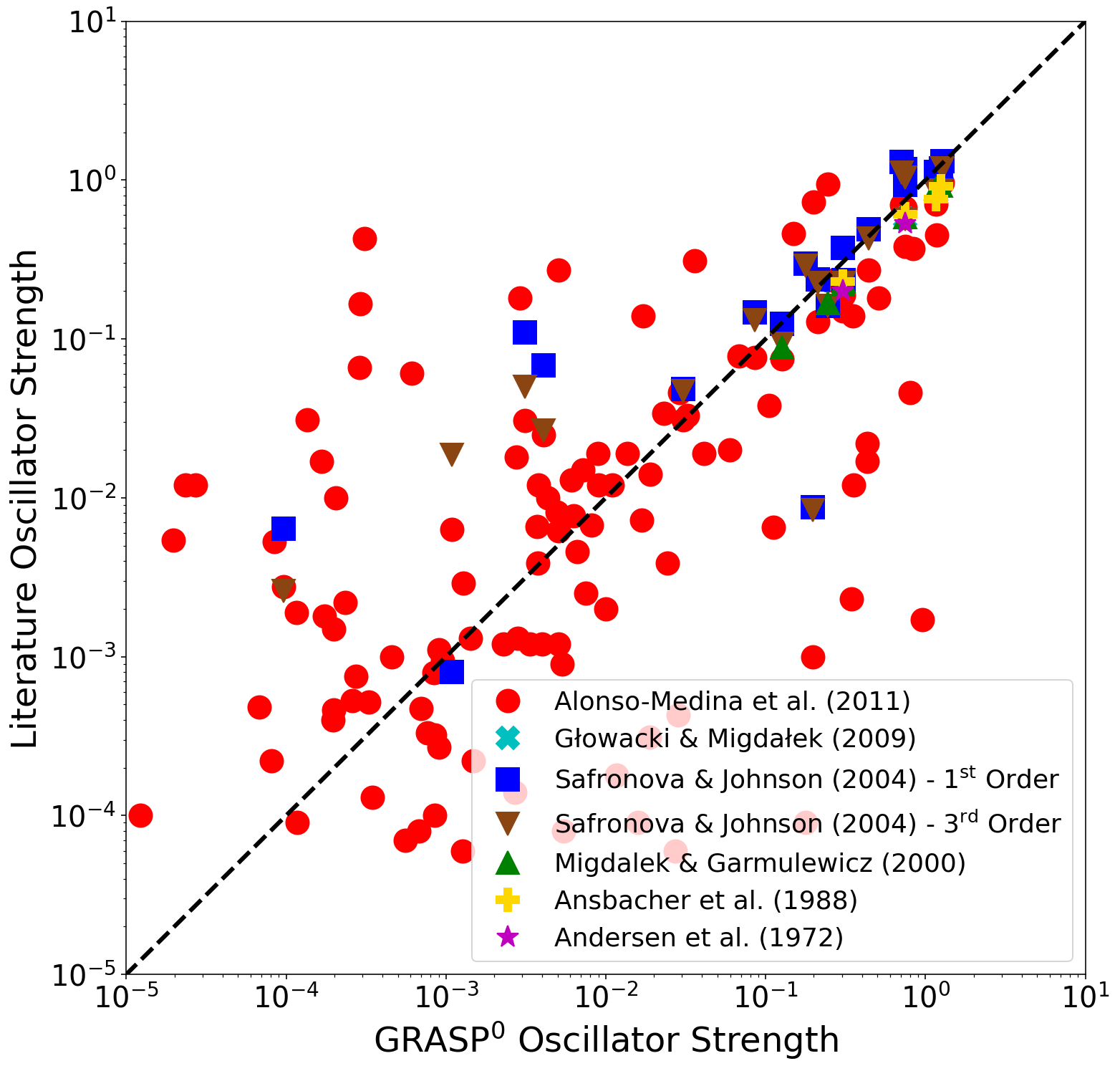}
\caption{A comparison of the oscillator strengths computed from our \ion{Pb}{iv} model with those experimentally determined in \protect\cite{Andersen_1972} and \protect\cite{Ansbacher_1988}, and theoretically calculated in \protect\cite{Migdalek_2000}, \protect\cite{Safronova_2004}, \protect\cite{Glowacki_2009} and \protect\cite{Alonso-Medina_2011}.}
\label{fig:PbIV_Oscillator_Strengths}
\end{figure}

To compare the oscillator strengths, we employ experimental values from \cite{Andersen_1972} and \cite{Ansbacher_1988} from beam-foil experiments for some of the lowest lying transitions.  There are also the oscillator strengths computed in \cite{Safronova_2004} by application of both a first and third order relativistic many body perturbation theory (MBPT).   We also reference the work of \cite{Migdalek_2000} and \cite{Glowacki_2009} who employ a Dirac-Fock approach with a polarisible ion-like core and configuration-interaction Dirac Fock approach respectively.  In addition, \cite{Alonso-Medina_2011} computed the oscillator strengths using the Cowan code previously discussed.  A graphical summary of these literature oscillator strengths as compared with our {\sc grasp} \ion{Pb}{iv} target is presented in Fig. \ref{fig:PbIV_Oscillator_Strengths}.  There is fairly good agreement with the oscillator strengths from our model to the other literature sources.  This is especially the case for the experimental oscillator strengths available.  In general, the stronger oscillator strengths have converged to the same magnitude.  Agreement on these is more important than for the weaker transitions, as these will correspond to the most prominent \ion{Pb}{iv} features observed in a spectra.  The agreement in the stronger A-values and oscillator strengths between our \ion{Pb}{iv} model and those in the literature provides confidence in the validity of our target structure.  

\subsection{\ion{Pb}{v} Structure}
\label{sec:Structure-PbV} 

Four times ionized lead has 78 electrons, and as such is a part of the platinum isoelectronic sequence.  It possesses a ground state of [Xe]4f$^{14}$5d$^{10}$($^1$S$_0$) and has an ionization potential of 5.06 Ryd \citep{Mack_1935}.  Our {\sc grasp}$^0$ Pb V target consists of 19 orbitals going up to $n=7$ and $l=3$.  The orbitals included were 1s, 2s, 2p, 3s, 3p, 3d, 4s, 4p, 4d, 4f, 5s, 5p, 5d, 6s, 6p, 6d, 7s, 7p and 7d.  The structure comprises of 14 unique configurations, as shown in Table \ref{tab:Pb_V_Structure}.  The primary configurations to include were the single electron promotion out of the 5d$^{10}$ orbital to the 7d orbital.  This resulted in a target structure comprising a total of 1,288 energy levels.

   \begin{table}
   
     \centering
        \caption[]{The configurations included in the wavefunction expansion for the {\sc grasp$^{0}$} structure of \ion{Pb}{v}.}
         \begin{tabular}{p{2cm} p{2cm } p{2cm}}
            \hline
            \noalign{\smallskip}
            \multicolumn{3}{l}{\textbf{\ion{Pb}{v} - 14 Configurations}} \\
            \noalign{\smallskip}
            \hline
            \noalign{\smallskip}
             5d$^{10}$         & 5d$^{9}$6s       & 5d$^{9}$6p       \\ \noalign{\smallskip}
             5d$^{9}$6d        & 5d$^{9}$7s       & 5d$^{9}$7p       \\ \noalign{\smallskip}
             5d$^{9}$7d        & 5d$^{8}$6s$^{2}$ & 5d$^{8}$6p$^{2}$ \\ \noalign{\smallskip}
             5d$^{8}$7d$^{2}$  & 5d$^{8}$6s7p     & 5d$^{8}$6p7p     \\ \noalign{\smallskip}
             5d$^{8}$6d7p      & 5d$^{8}$7s7p     &                  \\ \noalign{\smallskip}
            \hline
            \noalign{\smallskip}
         \end{tabular}

        \label{tab:Pb_V_Structure}
   \end{table}

There were 45 levels present in the NIST database for \ion{Pb}{v}, with the experimental energies determined in \cite{Gutmann_1969}, \cite{Joshi_1990} \& \cite{Wyart_1992}.  We compared the lowest lying 44 levels with their equivalent values in the target structure.  The first 20 energy levels, along with their NIST counterparts, where available, are shown in Table \ref{tab:Pb_V-Energy-Levels}.  Where a comparison is available, there is good agreement in the NIST database when compared to our \ion{Pb}{v} target, with the energy differences being $\leq$ 0.07 Ryd.  The average relative percentage difference between these 44 energy levels was found to be 0.604\%.

\begin{table}
  \caption{A comparison of the first 20 energy levels in our \ion{Pb}{v} target with their equivalent values in NIST.   The average relative percentage difference between all the shifted energy levels is 0.604\%. The NIST energies were obtained from \protect\cite{Joshi_1990}.}
  \centering
    \begin{tabular}{ccccc} \hline
            \textbf{Level} & \textbf{Config.}      & \TableNewLine{\textbf{NIST} \\ \textbf{/ Ryd}} & \TableNewLine{\textbf{GRASP$^0$} \\ \textbf{/ Ryd}} & \TableNewLine{\textbf{Percentage} \\ \textbf{Error / \%}} \\ \hline
 \noalign{\smallskip}
    1  & 5d$^{10}$  ($^1$S$_0$)               &  0.00000 & 0.00000 &             \\ \noalign{\smallskip}
    2  & 5d$^{9}$6s ($^3$D$_3)$               &  1.00941 & 1.06524 & 5.531           \\ \noalign{\smallskip}
    3  & 5d$^{9}$6s ($^3$D$_2)$               &  1.04530 & 1.10215 & 5.438           \\ \noalign{\smallskip}
    4  & 5d$^{9}$6s ($^3$D$_1)$               &  1.20939 & 1.25381 & 3.673           \\ \noalign{\smallskip}
    5  & 5d$^{9}$6s ($^1$D$_2)$               &  1.23933 & 1.28568 & 3.740           \\ \noalign{\smallskip}
    6  & 5d$^{9}$6p ($^3$P$_2^{\circ}$)       &  1.77521 & 1.79719 & 1.238           \\ \noalign{\smallskip}
    7  & 5d$^{9}$6p ($^3$F$_3^{\circ}$)       &  1.79643 & 1.82293 & 1.475           \\ \noalign{\smallskip}
    8  & 5d$^{9}$6p ($^3$F$_2^{\circ}$)       &  1.97810 & 1.99077 & 0.640           \\ \noalign{\smallskip}
    9  & 5d$^{9}$6p ($^3$P$_1^{\circ}$)       &  2.00017 & 2.00711 & 0.347           \\ \noalign{\smallskip}
    10 & 5d$^{9}$6p ($^3$F$_4^{\circ}$)       &  2.01453 & 2.01538 & 0.042           \\ \noalign{\smallskip}
    11 & 5d$^{9}$6p ($^1$D$_2^{\circ}$)       &  2.04044 & 2.05719 & 0.821           \\ \noalign{\smallskip}
    12 & 5d$^{9}$6p ($^3$D$_3^{\circ}$)       &  2.06413 & 2.07696 & 0.622           \\ \noalign{\smallskip}
    13 & 5d$^{9}$6p ($^1$P$_1^{\circ}$)       &  2.07623 & 2.09266 & 0.791           \\ \noalign{\smallskip}
    14 & 5d$^{8}$6s$^{2}$ ($^3$F$_4$)         &          & 2.09937 &                  \\ \noalign{\smallskip}
    15 & 5d$^{9}$6p ($^3$P$_0^{\circ}$)       &  2.16576 & 2.16557 & 0.008 \\ \noalign{\smallskip}
    16 & 5d$^{9}$6s$^{2}$\,($^1$D$_2$)        &          & 2.20584 &                  \\ \noalign{\smallskip}
    17 & 5d$^{9}$6p ($^3$F$_3^{\circ}$)       &  2.22954 & 2.22425 & 0.238 \\ \noalign{\smallskip}
    18 & 5d$^{9}$6p ($^3$D$_1^{\circ}$)       &  2.23515 & 2.23349 & 0.074 \\ \noalign{\smallskip}
    19 & 5d$^{9}$6p ($^3$D$_2^{\circ}$)       &  2.25628 & 2.26024 & 0.175           \\ \noalign{\smallskip}
    20 & 5d$^{9}$6s$^{2}$ ($^3$F$_3$)         &          & 2.27625 &                  \\ \noalign{\smallskip}
    \hline
    \noalign{\smallskip}
    \end{tabular}%

    \label{tab:Pb_V-Energy-Levels}%
\end{table}

\begin{figure}
\centering
\includegraphics[width=0.9\columnwidth]{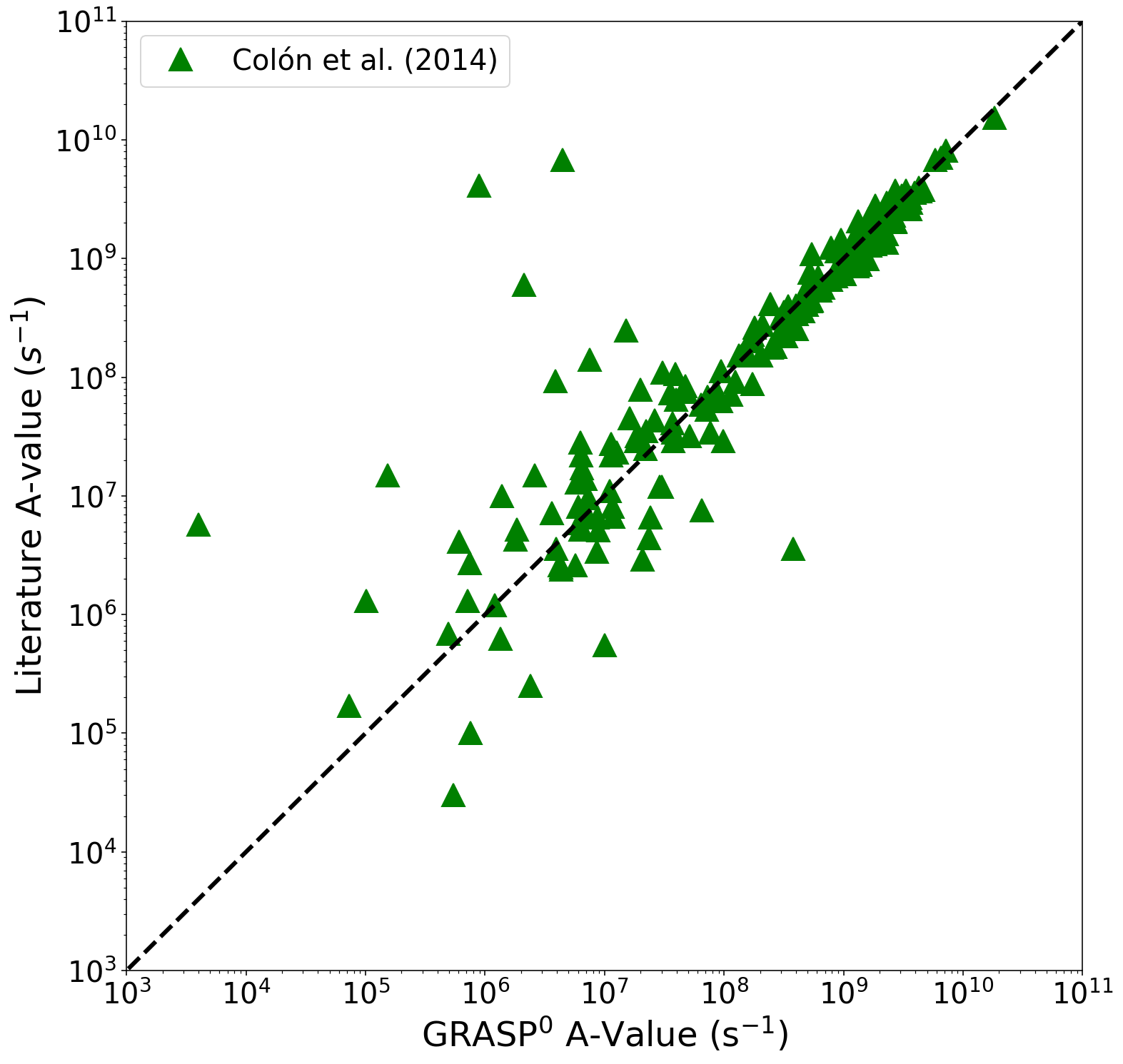}
\caption{A comparison between the A-values from our \ion{Pb}{v} model with those calculated in \protect\cite{Colon_2014}.}
\label{fig:PbV_A-values}
\end{figure}

We now present a comparison of the A-values generated from our \ion{Pb}{v} model to those calculated in \cite{Colon_2014} using the computer codes described in \cite{Cowan_1981}.  We considered only the A-values from our {\sc grasp$^0$} calculation where it was possible to spectroscopically shift both the upper and lower levels of the corresponding transition to NIST values.  The final comparison is illustrated in Fig. \ref{fig:PbV_A-values}, which consists of 190 A-values in total.  The A-values from our \ion{Pb}{v} model are overall in very good agreement with those reported in \cite{Colon_2014}, with the vast majority of them lying within the same order of magnitude.  This is further assurance that our \ion{Pb}{v} target structure is reliable.  


\subsection{\ion{Pb}{vi} Structure}
\label{sec:Structure-PbVI}

There are 77 electrons present in five times ionized lead, and as such it is a part of the iridium isoelectronic sequence.  It has a ground configuration of [Xe]4f$^{14}$5d$^{9}$($^2$D$_{5/2}$), and was determined in \cite{Rodrigues_2004} to have an ionization potential of 6.10 Ryd.  Our {\sc grasp}$^{0}$ model of \ion{Pb}{vi} consists of 17 orbitals going up to $n$=6 and $l$=3.  The orbitals included were 1s, 2s, 2p, 3s, 3p, 3d, 4s, 4p 4f, 5s, 5p, 5d, 5f, 6s, 6p and 6d.  We used 13 configurations when constructing the target, resulting in it having a total of 2,566 levels.  These included the single electron promotion out of the 5d$^9$ subshell, as well as various single and double promotions out the 5s$^2$ and 5p$^6$ orbitals to allow convergence.  The specific configurations included are shown in Table \ref{tab:Pb_VI_Structure}.  

   \begin{table}
   
     \centering
        \caption[]{The configurations included in the wavefunction expansion for the {\sc grasp$^{0}$} structure of \ion{Pb}{vi}.}
         \begin{tabular}{p{2cm} p{2cm } p{2cm}}
            \hline
            \noalign{\smallskip}
            \multicolumn{3}{l}{\textbf{\ion{Pb}{vi} - 13 Configurations}} \\
            \noalign{\smallskip}
            \hline
            \noalign{\smallskip}
              5s$^{2}$5p$^{6}$5d$^{9}$          & 5s$^{2}$5p$^{6}$5d$^{8}$5f    & 5s$^{2}$5p$^{6}$5d$^{8}$6s        \\ \noalign{\smallskip}
              5s$^{2}$5p$^{6}$5d$^{8}$6p        & 5s$^{2}$5p$^{6}$5d$^{8}$6d    & 5s$^{2}$5p$^{6}$5d$^{7}$6s$^{2}$  \\ \noalign{\smallskip}
              5s$^{2}$5p$^{6}$5d$^{7}$6d$^{2}$  & 5s$^{2}$5p$^{5}$5d$^{10}$     & 5s$^{2}$5p$^{5}$5d$^{9}$6s        \\ \noalign{\smallskip}
              5s$^{2}$5p$^{4}$5d$^{9}$5f$^{2}$  & 5s5p$^{6}$5d$^{10}$           & 5s5p$^{6}$5d$^{9}$6s              \\ \noalign{\smallskip}
              5p$^{6}$5d$^{9}$5f$^{2}$          &                               &                                   \\ \noalign{\smallskip}

            \hline
            \noalign{\smallskip}
         \end{tabular}
        
        \label{tab:Pb_VI_Structure}
   \end{table}

We compared the energy levels arising from our structure with those experimentally determined in \cite{Raassen_1990_PbVI}.  This allowed a comparison of 60 of the energy levels in our target structure.  The 20 lowest lying energy levels in our target, along with their equivalent values from the literature if present, are highlighted in Table \ref{tab:PbVI-Energy-Levels}.  There is good agreement between the energy levels generated from our structure to those reported in \cite{Raassen_1990_PbVI}, where differences are $\leq 0.04$Ryd.  The average relative percentage difference arising from the model was found to be 1.205\%, and no level was found to deviate from the structure by 3\%.

\begin{table}
  \caption{A comparison of the first 20 levels in our \ion{Pb}{vi} structure with those experimentally observed in \protect\cite{Raassen_1990_PbVI}.  These are referred to in the table as (RAA. 1990).  The average relative percentage difference between all of the shifted energy levels is 1.205\%.}
  \centering
\resizebox{0.99\columnwidth}{!}{
    \begin{tabular}{ccccc} \hline
            \textbf{Level} & \textbf{Config.}      & \TableNewLine{\textbf{RAA. 1990} \\ \textbf{/ Ryd}} & \TableNewLine{\textbf{GRASP$^0$} \\ \textbf{/ Ryd}} & \TableNewLine{\textbf{Percentage} \\ \textbf{Error / \%}} \\ \hline
 \noalign{\smallskip}
    1     &  5d$^9$  ($^2$D$_{5/2}$)             & 0.00000 & 0.00000 &           \\ \noalign{\smallskip}
    2     &  5d$^9$  ($^2$D$_{3/2}$)             &         & 0.19200 &           \\ \noalign{\smallskip}
    3     &  5d$^8$6s ($^4$F$_{9/2}$)            & 1.20039 & 1.19167 & 0.726     \\ \noalign{\smallskip}
    4     &  5d$^8$6s ($^4$F$_{7/2}$)            & 1.26453 & 1.25975 & 0.378     \\ \noalign{\smallskip}
    5     &  5d$^8$6s ($^2$D$_{5/2}$)            & 1.30146 & 1.31794 & 1.266     \\ \noalign{\smallskip}
    6     &  5d$^8$6s ($^2$D$_{3/2}$)            & 1.32375 & 1.33431 & 0.798     \\ \noalign{\smallskip}
    7     &  5d$^8$6s ($^4$F$_{5/2}$)            & 1.41898 & 1.40447 & 1.023     \\ \noalign{\smallskip}
    8     &  5d$^8$6s ($^2$F$_{7/2}$)            & 1.42605 & 1.41265 & 0.940     \\ \noalign{\smallskip}
    9     &  5d$^8$6s ($^4$P$_{1/2}$)            & 1.44676 & 1.48341 & 2.534     \\ \noalign{\smallskip}
    10    &  5d$^8$6s ($^4$P$_{3/2}$)            & 1.48617 & 1.49413 & 0.536     \\ \noalign{\smallskip}
    11    &  5d$^8$6s ($^2$F$_{5/2}$)            & 1.50534 & 1.50894 & 0.239     \\ \noalign{\smallskip}
    12    &  5d$^8$6s ($^2$P$_{3/2}$)            & 1.54104 & 1.55685 & 1.026     \\ \noalign{\smallskip}
    13    &  5d$^8$6s ($^2$G$_{9/2}$)            & 1.57044 & 1.59371 & 1.482     \\ \noalign{\smallskip}
    14    &  5d$^8$6s ($^2$G$_{7/2}$)            & 1.57974 & 1.60157 & 1.382     \\ \noalign{\smallskip}
    15    &  5d$^8$6s ($^4$F$_{9/2}$)            & 1.58297 & 1.61491 & 2.018     \\ \noalign{\smallskip}
    16    &  5d$^8$6s ($^2$P$_{1/2}$)            & 1.68373 & 1.68126 & 0.147     \\ \noalign{\smallskip}
    17    &  5d$^8$6s ($^2$D$_{3/2}$)            & 1.69633 & 1.68298 & 0.787     \\ \noalign{\smallskip}
    18    &  5d$^8$6s ($^2$S$_{1/2}$)            &         & 2.00238 &            \\ \noalign{\smallskip}
    19    &  5d$^8$6p ($^4$D$_{7/2}^{\circ}$)    & 2.04674 & 1.98921 & 2.811     \\ \noalign{\smallskip}
    20    &  5d$^8$6p ($^2$G$_{9/2}^{\circ}$)    & 2.07558 & 2.03059 & 2.168     \\ \noalign{\smallskip}
    \hline
    \noalign{\smallskip}
    \end{tabular}%
}
    \label{tab:PbVI-Energy-Levels}%
\end{table}

A comparison with the A-values arising from our target structure was made with the work of \cite{Uylings_1995}, who calculated five A-values corresponding to selected electric dipole transitions between the 5d$^9$ and 5d$^8$6p configurations using orthogonal operators with relativistic MCDF wave functions.  In addition, they performed the calculation by considering the 5d$^8$6s configuration positioned either inside (full diagonalization) or outside (perturbation) the model space.  We compared the corresponding A-values from our {\sc grasp$^0$} model to the A-values calculated from both of these methods, as shown in Table \ref{tab:PbVI-A-values}.  Our A-values are in good agreement to those calculated by both methods in \cite{Uylings_1995}, with both sets of A-values lying within the same order of magnitude.

\begin{table}
    \caption{Comparison between Einstein A-values obtained from our \ion{Pb}{vi} {\sc grasp}$^0$ target and those determined in \protect\cite{Uylings_1995} ([1]- Diagonalisation Method and [2]- Perturbation Method).  The Index column refers to the energy levels displayed in Table \ref{tab:PbIV-Energy-Levels} and in the accompanying {\sc topbase} file.}    
  \centering
    \begin{tabular}{cccc} \hline
            \textbf{Index} & \TableNewLine{\textbf{Wavelength} \\ \textbf{ / nm}}      & \TableNewLine{\textbf{GRASP$^0$} \\ \textbf{A-value / s$^{-1}$} } & \TableNewLine{\textbf{Literature} \\ \textbf{A-value / s$^{-1}$}}  \\ \hline
 \noalign{\smallskip}
    1 - 21                & 427.69  & 9.86E+07 & 1.89E+08$^{[1]}$ \\ \noalign{\smallskip}
    & & & 1.78E+08$^{[2]}$ \\ \noalign{\smallskip}
    1 - 65                & 323.52  & 4.39E+07 & 2.22E+08$^{[1]}$ \\ \noalign{\smallskip}
    & & & 1.27E+07$^{[2]}$ \\ \noalign{\smallskip}
    2 - 22     & 470.40  & 1.01E+07 & 1.32E+08$^{[1]}$ \\ \noalign{\smallskip}
    & & & 9.78E+07$^{[2]}$ \\ \noalign{\smallskip}
    2 - 28                & 425.65  & 8.28E+06 & 9.02E+06$^{[1]}$ \\ \noalign{\smallskip}
    & & & 4.35E+06$^{[2]}$ \\ \noalign{\smallskip}
    2 - 38                & 400.01  & 1.16E+08 & 4.70E+07$^{[1]}$ \\ \noalign{\smallskip}
    & & & 7.33E+07$^{[2]}$ \\ \noalign{\smallskip}

    \hline
    \noalign{\smallskip}
    \end{tabular}%
\label{tab:PbVI-A-values}%
\end{table}

\subsection{\ion{Pb}{vii} Structure}
\label{sec:Structure-PbVII}

Six times ionized lead is part of the osmium isoelectronic sequence, and as such, contains 76 electrons in its structure.  It has a ground state configuration of [Xe]4f$^{14}$5d$^7$ ($^3$F$_4$), and was calculated in \cite{Rodrigues_2004} to have an ionization potential of 7.35 Ryd.  Our \ion{Pb}{vii} target consists of 19 orbitals extending to $n$=7 and $l$=3.  The orbitals were 1s, 2s, 2p, 3s, 3p, 3d, 4s, 4p, 4d, 4f, 5s, 5p, 5d, 6s, 6p, 6d, 7s, 7p and 7d.  The structure consists of 10 unique configurations listed in Table \ref{tab:Pb_VII_Structure}, resulting in the target having 1,679 distinct energy levels in total.  These included the single electron promotion out of the 5d$^8$ orbital, with additional 5d$^6$6s $nl$ promotions to achieve convergence.

   \begin{table}
   
     \centering
        \caption[]{The configurations included in the wavefunction expansion for the {\sc grasp$^{0}$} structure of \ion{Pb}{vii}.}
         \begin{tabular}{p{2cm} p{2cm } p{2cm}}
            \hline
            \noalign{\smallskip}
            \multicolumn{3}{l}{\textbf{\ion{Pb}{vii} - 10 Configurations}} \\
            \noalign{\smallskip}
            \hline
            \noalign{\smallskip}
             5d$^8$         & 5d$^7$6s     & 5d$^7$6p        \\ \noalign{\smallskip}
             5d$^7$6d       & 5d$^7$7s     & 5d$^7$7p        \\ \noalign{\smallskip}
             5d$^7$7d       & 5d$^6$6s6p   & 5d$^6$6s6d      \\ \noalign{\smallskip}
             5d$^6$6s7s     &              &                 \\ \noalign{\smallskip}
            \hline
            \noalign{\smallskip}
         \end{tabular}

        \label{tab:Pb_VII_Structure}
   \end{table}

As a test of the validity of our target structure, we compared our energy levels to those experimentally measured in \cite{Raassen_1990_PbVII} \& \cite{Raassen_1994}. We aligned 120 out of the 157 energy levels from the 5d$^8$, 5d$^7$6s and 5d$^7$6p configurations available in the literature with their equivalent values in our target. A sample is displayed in Table \ref{tab:PbVII-Energy-Levels}.  Across the 120 levels where a comparison was possible, there is very good agreement in the energy levels arising from our \ion{Pb}{vii} target and those determined experimentally.  The energy levels typically deviated by $\leq 0.04$Ryd.  The average relative percentage difference was $-$0.135\%, though it should be noted that this is inflated by the very large percentage deviation ($>$24\%) arsing from Level 2.  The absolute energy difference in Level 2 is similar to those seen in the other levels.

\begin{table}
  \caption{A comparison of the first 20 levels in our \ion{Pb}{vii} structure with those experimentally observed in [1] - \protect\cite{Raassen_1990_PbVII} \& [2] - \protect\cite{Raassen_1994}.  The average relative percentage difference between all of the shifted energy levels is -0.135\%.}
  \centering
\resizebox{0.99\columnwidth}{!}{
    \begin{tabular}{ccccc} \hline
            \textbf{Level} & \textbf{Config.}      & \TableNewLine{\textbf{Literature} \\ \textbf{/ Ryd}} & \TableNewLine{\textbf{GRASP$^0$} \\ \textbf{/ Ryd}} & \TableNewLine{\textbf{Percentage} \\ \textbf{Error / \%}} \\ \hline
 \noalign{\smallskip}
    1  & 5d$^8$ ($^3$F$_4$)       & 0.00000$^{[1]}$  &  0.00000 &  \\ \noalign{\smallskip}
    2  & 5d$^8$ ($^1$D$_2$)       & 0.08030$^{[1]}$  &  0.09986 & 24.366\\ \noalign{\smallskip}
    3  & 5d$^8$ ($^3$F$_3$)       & 0.19332$^{[1]}$  &  0.18427 & \phantom{-}4.684\\ \noalign{\smallskip}
    4  & 5d$^8$ ($^3$P$_0$)       & 0.22989$^{[1]}$  &  0.26807 & 16.604\\ \noalign{\smallskip}
    5  & 5d$^8$ ($^3$P$_2$)       & 0.26746$^{[1]}$  &  0.27596 & \phantom{-}3.177\\ \noalign{\smallskip}
    6  & 5d$^8$ ($^3$P$_1$)       & 0.30695$^{[1]}$  &  0.33112 & \phantom{-}7.875\\ \noalign{\smallskip}
    7  & 5d$^8$ ($^1$G$_4$)       & 0.34572$^{[1]}$  &  0.36975 & \phantom{-}6.951\\ \noalign{\smallskip}
    8  & 5d$^8$ ($^1$D$_2$)       & 0.45781$^{[1]}$  &  0.45752 & \phantom{-}0.062\\ \noalign{\smallskip}
    9  & 5d$^8$ ($^1$S$_0$)       &           &  0.77583 &  \\ \noalign{\smallskip}
    10 & 5d$^7$6s ($^5$F$_5$)     & 1.46430$^{[2]}$  &  1.42904 & \phantom{-}2.408\\ \noalign{\smallskip}
    11 & 5d$^7$6s ($^5$F$_4$)     & 1.54774$^{[2]}$  &  1.52176 & \phantom{-}1.679\\ \noalign{\smallskip}
    12 & 5d$^7$6s ($^5$F$_3$)     & 1.62824$^{[2]}$  &  1.60631 & \phantom{-}1.347\\ \noalign{\smallskip}
    13 & 5d$^7$6s ($^3$P$_2$)     & 1.60996$^{[2]}$  &  1.60876 & \phantom{-}0.075\\ \noalign{\smallskip}
    14 & 5d$^7$6s ($^5$F$_1$)     &           &  1.62964 &  \\ \noalign{\smallskip}
    15 & 5d$^7$6s ($^5$F$_2$)     & 1.66911$^{[2]}$  &  1.65198 & \phantom{-}1.027\\ \noalign{\smallskip}
    16 & 5d$^7$6s ($^3$F$_4$)     & 1.68288$^{[2]}$  &  1.65485 & \phantom{-}1.665\\ \noalign{\smallskip}
    17 & 5d$^7$6s ($^5$P$_3$)     & 1.70474$^{[2]}$  &  1.68246 & \phantom{-}1.307\\ \noalign{\smallskip}
    18 & 5d$^7$6s ($^5$P$_1$)     &           &  1.73753 &  \\ \noalign{\smallskip}
    19 & 5d$^7$6s ($^3$G$_5$)     & 1.77242$^{[2]}$  &  1.75597 & \phantom{-}0.928\\ \noalign{\smallskip}
    20 & 5d$^7$6s ($^3$F$_3$)     & 1.78351$^{[2]}$  &  1.76974 & \phantom{-}0.772\\ \noalign{\smallskip}
    \hline
    \noalign{\smallskip}
    \end{tabular}%
}
    \label{tab:PbVII-Energy-Levels}%
\end{table}

\begin{figure}
\centering

\includegraphics[width=0.9\columnwidth]{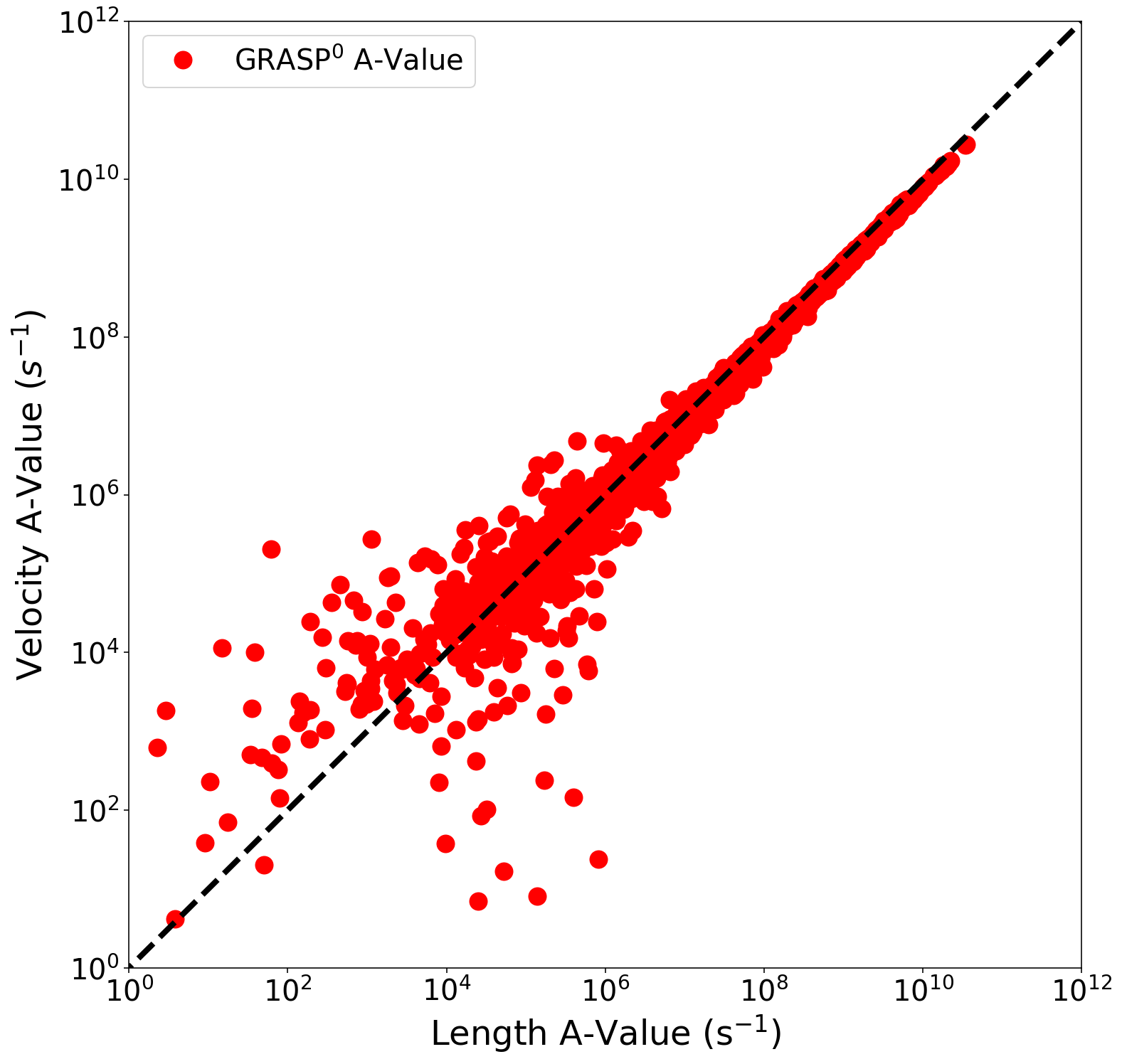}
\caption{A comparison of the A-values corresponding to electric dipole transitions arising from our \ion{Pb}{vii} model, considering both length and velocity gauge treatments.}
\label{fig:PbVII_A-values}
\end{figure}

A search of the literature did not discover any supplementary A-values or oscillator strengths for \ion{Pb}{vii} that we could use for comparison to provide additional validation of our target structure.  However, to provide further confidence that our resulting A-values had stabilised to their final values, we recalculated the A-values by consideration of the length and velocity gauges.  We performed this for all of the electric dipole transitions arising between the lowest lying 157 levels in our target structure.  This resulted in a comparison consisting of 2,548 A-values in total.  Fig. \ref{fig:PbVII_A-values} summarises this comparison.  It can be seen that the A-values show a discrepancy in their magnitudes at lower magnitudes between the length and velocity gauge, but tend to converge to the same value for stronger transitions.  It is more important that the stronger A-values have converged to approximately the same value, as these will correlate to the most frequently occurring transitions, and hence are the lines which will appear the strongest in a \ion{Pb}{vii} absorption spectrum.  It should be noted that this test was also performed for the other three Pb models described, and each one followed the same trends in their respective A-values, as illustrated in Fig. \ref{fig:PbVII_A-values}.

\begin{figure*}
\centering

\includegraphics[scale=0.35]{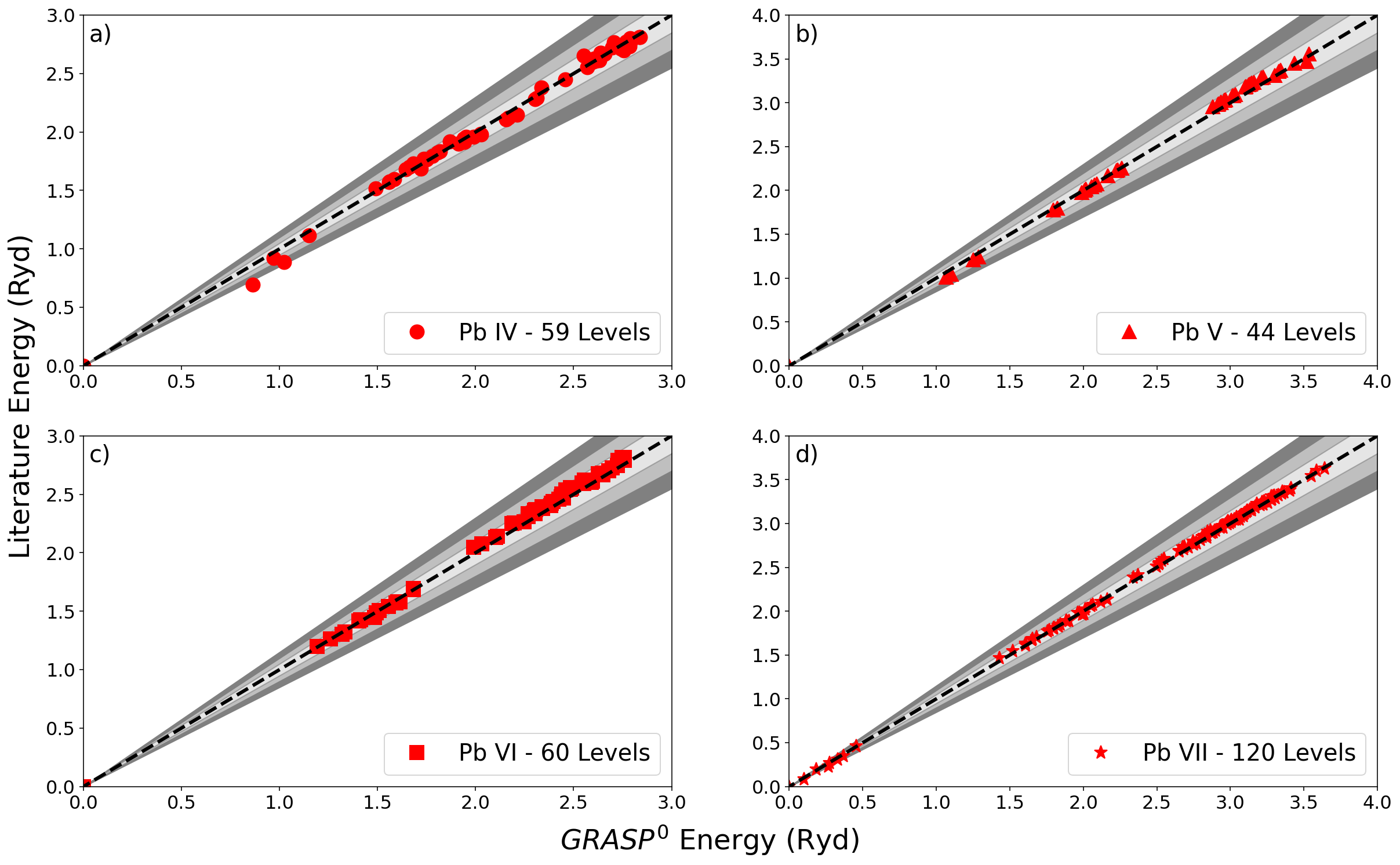}
\caption{Comparison between the energy levels derived from our {\sc grasp$^0$} models of a) \ion{Pb}{iv}, b) \ion{Pb}{v}, c) \ion{Pb}{vi} \& d) \ion{Pb}{vii}, and their equivalent values in NIST and other literature sources.  The experimental sources for each species were: \ion{Pb}{iv} \protect\citep[][]{Moore_1971,Gutmann_1973,Raassen_1991}, \ion{Pb}{v} \protect\citep[][]{Gutmann_1969,Joshi_1990,Wyart_1992}, \ion{Pb}{vi} \protect\citep[][]{Raassen_1990_PbVI} and \ion{Pb}{vii} \citep[][]{Raassen_1990_PbVII,Raassen_1994}.  The shaded gray regions indicate a $\leq$5\% (light), $\leq$10\% (medium) and $\leq$15\% (dark) deviation from the black dashed line of equality.}
\label{fig:Pb_Energy_Comparison}
\end{figure*}

In Fig. \ref{fig:PbIV_A-values}, Fig. \ref{fig:PbV_A-values}, Fig. \ref{fig:PbVII_A-values} and Table \ref{tab:PbVI-A-values}, we have shown the conformity between the A-values calculated in this work and all data currently available in the literature.  For completeness, we present in Fig. \ref{fig:Pb_Energy_Comparison} similar graphical evidence for the accuracy of the energy levels for each Pb species considered.  
We note that to allow for more precise identification of lines in an observational spectra, the collisional calculations discussed in the next section had, where possible, the energy levels systematically shifted to their corresponding spectroscopic positions.

\section{Photoionization Calculations}
\label{sec:PhotoionizationCalculations}

\subsection{$R$-Matrix Methodology}

The $R$-matrix approach introduced by \cite{Wigner_1947} and expanded in \cite{Burke_1968} \& \cite{Burke_1971} is considered a very effective technique for calculating the collisional interactions within high density matrix systems.  A brief summary, along with its specific application for photoionization calculations, is described below, but a more comprehensive derivation of the technique is provided in \cite{Burke_2011}.

For a specified $N$-electron target system, the $R$-matrix calculation is divided into spherical internal and external regions positioned at the center of the target.  The incident and subsequently scattered electron is represented by two different wave functions within the internal and external regions and is continuous across the boundary.  In the internal region, the exchange forces and correlation effects arising from the interactions between bound electrons provide a significant contribution to the resulting wave function.  Within the external region, the effects from electron interactions are minimal, and the corresponding wave function is approximated as the electron acting under the long range potential exerted by the target nuclei.  The boundary between the inner and outer regions is equated to the maximum extent of the most diffuse orbital in the target structure.  The $R$-matrix calculation itself ($R_{ij}$) of the total $N+1$ electron system is defined as

\begin{equation}
    R_{ij} = \frac{1}{2a}\sum^{N+1}_{k}\frac{\omega_{ik}(a)\omega_{jk}(a)}{E_k^{N+1}-E}
\label{R-Matrix}    
\end{equation}

where $a$ is the $R$-matrix boundary, $E_k^{N+1}$ is the eigenenergies of the $N+1$ Hamiltonian, $E$ is the energy of the incident electron, and $\omega_{ik/jk}$ are the surface amplitudes.  An extension to the $R$-matrix approach for photoionization calculations was developed in \cite{Burke_1975}, where in addition, the dipole matrices between all E1-connected symmetries are also determined across all of the unique $J\pi$ symmetries included in the $N+1$ electron system.  These are used to calculate the positions of the bound states present in the $N+1$ system from which photoionization may occur.  The bound states are identified in batches of unique $J\pi$ symmetries.  The subsequent photoionization cross sections from an identified level is determined through the application of the Wiger-Eckart Theorem on a dipole matrix between the initial and $R$-matrix basis states.  Averaging over polarisation states and integrating over the possible ejection angles yields the finalized photoionization cross section ($\sigma$) between an initial ($i$) and final ($f$) state as:

\begin{equation}
    \sigma_{i\rightarrow f} = \frac{8\pi^{2}\alpha a^2\omega}{3(2J_i + 1)} \sum_{f}|\langle \Psi_f^{J\pi^-}||D||\Psi_i^{J\pi}\rangle|^2
\end{equation}

where $\alpha$ is the fine structure constant, $\omega$ is the incident photon energy, $D$ is the Electric dipole operator where $D=-e\mathbf{r}$, $\Psi^{J\pi}_i$ is the wave function of the initial state, and $\Psi_f^{J\pi}$ is the wave function of the $R$-matrix basis states.  The cross sections are calculated across a wide range of different incident photon energies, as determined through a user defined photon-energy mesh grid.  We aimed to create a mesh grid containing a substantial number of points to both map any possible resonance structure at lower incident energies, as well as to extend to higher energies to ensure the signature gradual decay for photoionization is present in the data set.  

It is important to note that these cross sections may be determined under a choice of different gauges.  There is the Babushkin (Length) Gauge, where a greater weight is prioritized on the outer components of the wave function \citep{Hibbert_1974}.  Conversely, there is the Coulomb (Velocity) Gauge, where the greater emphasis is placed on the inner part of the wave function \citep{Papoulia_2019}.  Under a quantum mechanical treatment, the $R$-matrix calculation will be gauge invariant and as such should not affect the final cross sections.  We repeated each calculation considering the Babushkin and Coulomb Gauges, and confirmed that the differences between the length and velocity gauges were minimal.  The photoionization cross sections accompanying this work were all determined using the Babushkin Gauge as other works suggest that it provides more accurate values at the lower, non-relativistic energies required for the study of astrophysical plasmas \citep[e.g.][]{Friedrich_2017,Rynkun_2022,Gaigalas_2024}.

We employed the Dirac Atomic $R$-matrix Code ({\sc darc}) to compute the collisional $R$-matrix calculations, which solves the Dirac Hamiltonian with a fully relativistic $jj$ coupled scattering calculation \citep[e.g.][]{Norrington_1987,Ballance_2004,Smyth_2019,FernandezMenchero_2020}.  The {\sc darc} codes may be obtained at \cite{Ballance_2025}.

We will now discuss the parameters employed for each of our $R$-matrix photoionization calculations, and show sample cross sections for \ion{Pb}{iii}, {\sc iv}, {\sc v} and {\sc vi} from both the ground and non-ground levels.  An extensive search of the literature was unsuccessful in finding other equivalent experimental or theoretical data to facilitate a comparison.  However, we can check for self-consistency by comparing how the cross sections vary over the different Pb charged states, and over different electronic configurations.  This confirms that the expected trends for photoionization in our data are followed.  We can also compare the energies of the bound states and confirm that the corresponding ionization potentials ($I_p$) align with those from NIST and other experimental sources.  A complete set of photoionization cross sections computed in this work, convoluted over a Gaussian function, is available in the accompanying {\sc topbase} files.   
 
\subsection{Photoionization of Pb  {\sc iii}}

The 691 energy levels present in our \ion{Pb}{iv} model were reduced to the lowest lying 100 for the close coupled calculations.  The 59 levels with an equivalent NIST energy were subsequently shifted to their corresponding literature values.  The remaining 41 levels were also shifted based on the mean percentage difference from the 59 matched pairs ( -0.741\%).  A continuum orbital basis of 25 was assigned to each dipole pair, with the $R$-matrix boundary between the inner and outer region set to 17.94au from the center of the target.  The scattering calculations were performed from $0 \leq J \leq 6$, resulting in 36 partial waves in total, with 18 unique dipole pairs.  The $N+1$ Hamiltonian peaked at 13422 $\times$ 13422, with the peak close coupled channel number being 534.

\begin{figure*}
\centering
\includegraphics[scale=0.35]{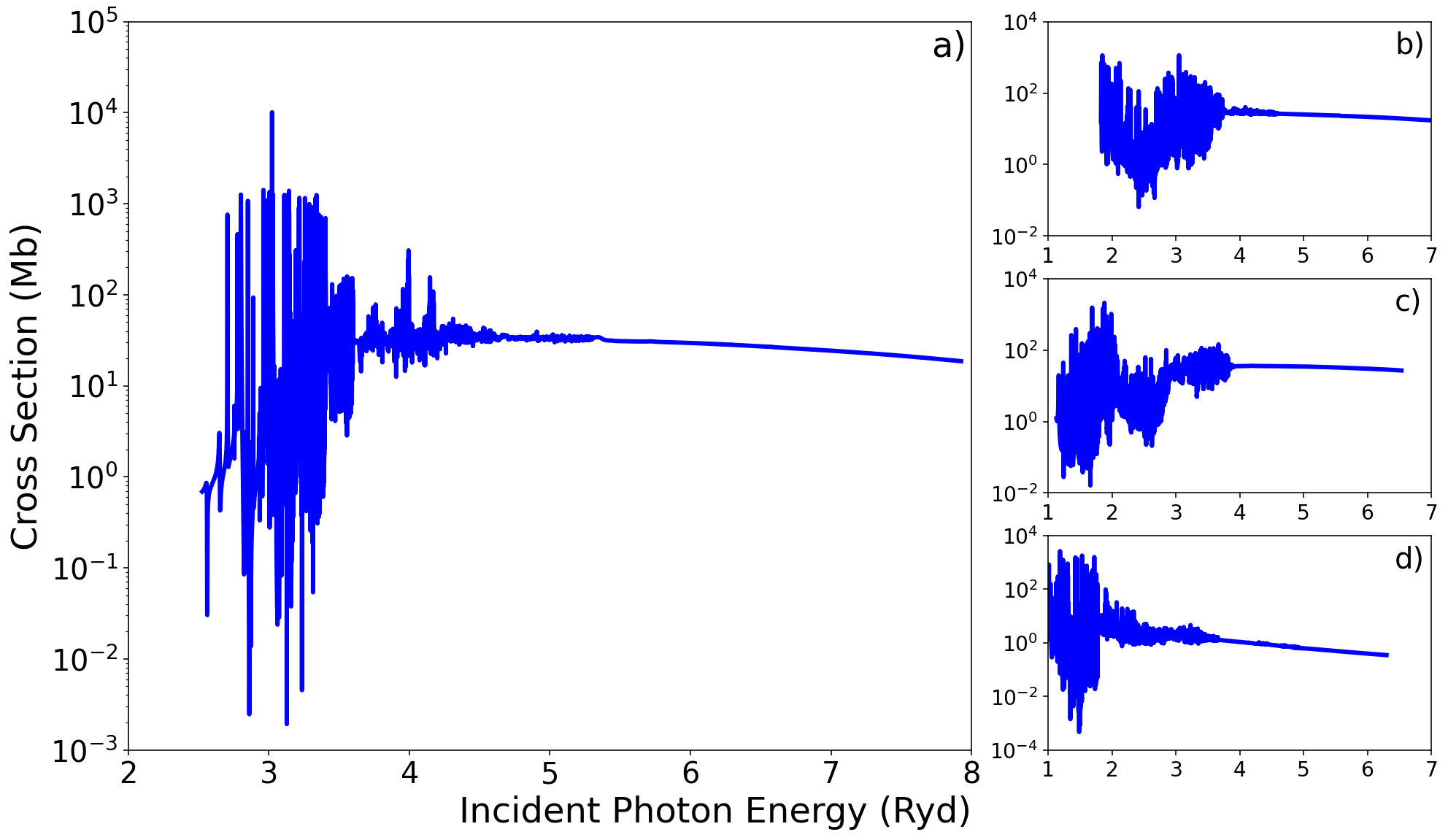}
\caption{Sample photoionization cross sections of \ion{Pb}{iii} for Levels a) - 1 (5d$^{10}$6s$^{2}$ ($^1$S$_0$)), b) - 4 (5$d^{10}$6s6p ($^3$P$_2^{\circ}$)), c) - 7 (5d$^{10}$6s7s ($^3$S$_1$)) and d) - 10 (5d$^9$6s$^2$6p ($^3$P$_2^{\circ}$)).  Cross sections for the other levels are included in the accompanying {\sc topbase} file.} 
\label{fig:PbIII_Cross_Sections}
\end{figure*}

The photon energy mesh grid was defined as having 120,000 points, starting at 0.005 Ryd and increasing in regular intervals of 5$\times$10$^{-6}$ Ryd.  This allowed the photoionization cross sections to be determined for energies up to 7.92 Ryd.  The cross sections from the ground and selected excited states are shown in Fig. \ref{fig:PbIII_Cross_Sections}.  We note that the photoionization cross sections from Level 10 - 5d$^{10}$6s$^2$6p ($^3P_2^{\circ}$), and other levels which share the same electronic configuration, are an order of magnitude lower to the other cross sections in this species.  This is to be expected, as this particular transition would occur less frequently from a level less likely to be significantly populated due to it originating from a inner orbital promotion.    The ionization potentials arising from our model are displayed in Table \ref{tab:PbIII-Ionization_Potential}.  The ionization potentials are in good agreement with the energies cited in NIST, with the differences in energy being $\leq$ 0.2 Ryd. 

\begin{table}
  \caption{The ionization potential arising from our {\sc darc} photoionization calculations for the first 10 levels in \ion{Pb}{iii}, compared with their equivalent values in NIST.  The ionization potential of \ion{Pb}{iii} is taken to be 2.347 Ryd (\protect\cite{Moore_1971}).  $I_p$ NIST was calculated from the works of \protect\cite{Moore_1971} \& \protect\cite{Martin_1972}. }
  \centering

    \begin{tabular}{cccc} \hline
            \textbf{Level} & \textbf{Config.}      & \TableNewLine{\textbf{$I_p$ NIST} \\ \textbf{/ Ryd}} & \TableNewLine{\textbf{$I_p$ DARC} \\ \textbf{/ Ryd}} \\ \hline
 \noalign{\smallskip}
    1  &  5d$^{10}$6$s^2$ ($^1$S$_0$)             & 2.347 & 2.481 \\ \noalign{\smallskip}
    2  &  5d$^{10}$6s6p ($^3$P$_0^{\circ}$)       & 1.797 & 1.946 \\ \noalign{\smallskip}
    3  &  5d$^{10}$6s6p ($^3$P$_1^{\circ}$)       & 1.761 & 1.910 \\ \noalign{\smallskip}
    4  &  5d$^{10}$6s6p ($^3$P$_2^{\circ}$)       & 1.628 & 1.782 \\ \noalign{\smallskip}
    5  &  5d$^{10}$6s6p ($^1$P$_1^{\circ}$)       & 1.479 & 1.595 \\ \noalign{\smallskip}
    6  &  5d$^{10}$6p$^{2}$ ($^3$P$_0$)           & 1.048 & 1.180 \\ \noalign{\smallskip}
    7  &  5d$^{10}$6s7s ($^3$S$_1$)               & 0.978 & 1.084 \\ \noalign{\smallskip}
    8  &  5d$^{10}$6s6d ($^1$D$_2$)               & 0.963 & 1.092 \\ \noalign{\smallskip}
    9  &  5d$^{10}$6s7s ($^1$S$_0$)               & 0.946 & 1.042 \\ \noalign{\smallskip}
    10 &  5d$^{9}$6s$^{2}$6p ($^3$P$_2^{\circ}$)  & 0.939 & 0.851 \\ \noalign{\smallskip}
    \hline
    \noalign{\smallskip}
    \end{tabular}%
    \label{tab:PbIII-Ionization_Potential}%
\end{table}

\subsection{Photoionization of \ion{Pb}{iv}}

The 1,288 energy levels present in our \ion{Pb}{v} target was cut back to the first 100 energy levels for the scattering calculations.  Of these, the 44 energy levels having a corresponding NIST value were spectroscopically shifted.  The remaining 56 energy levels were shifted based on the average percentage difference (0.604\%) arising from the other energy levels.  A continuum orbital of 22 was set for each dipole pair, with the $R$-matrix boundary between the inner and outer region defined at 14.72au from the target center.  The scattering calculations were repeated for all dipole pairs where 0.5 $\leq J \leq$ 5.5, resulting in 32 partial waves across 16 unique dipole pairs.  The $N+1$ Hamiltonians peaked at 12423 $\times$ 12423, with the peak close channel number extending to 563.

\begin{figure*}
\centering
\includegraphics[scale=0.35]{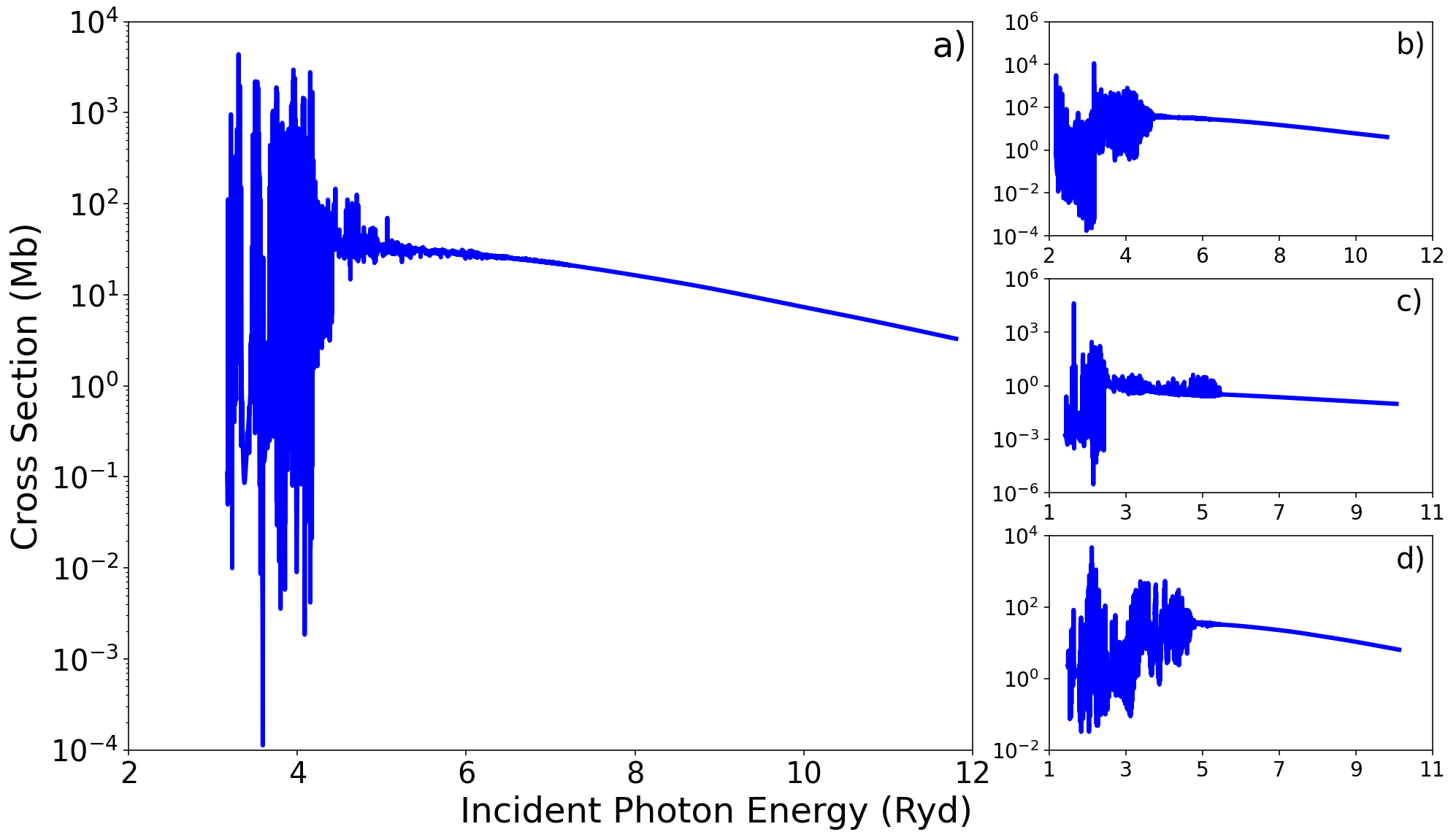}
\caption{Sample photoionization cross sections of \ion{Pb}{iv} for Levels a) - 1 (5d$^{10}$6s ($^2$S$_{1/2}$)), b) - 4 (5d$^{9}$6s$^2$ ($^2$D$_{5/2}$)), c) - 7 (5d$^{9}$6s6p ($^4$F$_{7/2}^{\circ}$)) and d) - 10 (5d$^{10}$6d ($^2$D$_{3/2}$)).  Cross sections for the other levels are included in the accompanying {\sc topbase} file.} 
\label{fig:PbIV_Cross_Sections}
\end{figure*}

A photon energy mesh grid consisting of 36,000 points was selected, starting at 1$\times$10$^{-5}$ Ryd and increasing in regular intervals by 1.5$\times$10$^{-5}$ Ryd.  This will cover photon energies of up to 11.8 Ryd.  Sample photoionization cross sections determined from this calculation are displayed in Fig. \ref{fig:PbIV_Cross_Sections}, with the ionization potentials arising from the first 10 levels of this model shown in Table \ref{tab:PbIV-Ionization_Potential}.  Again, we note the expected result that the levels originating from inner shell promotions exhibit significantly smaller photoionization cross sections, as seen with Level 7 - 5d$^9$6s6p ($^4F_{7/2}^{\circ}$).   There is fairly good agreement in the ionization potentials arising from our {\sc darc} calculation to those derived from NIST, where the differences are typically $\leq$ 0.1 Ryd.

\begin{table}
  \caption{The ionization potential arising from our {\sc darc} photoionization calculations for the first 10 levels in \ion{Pb}{iv}, compared with their equivalent values in NIST.  The ionization potential of \ion{Pb}{iv} is taken to be 3.111 Ryd (\protect\cite{Hanni_2010}).  $I_p$ NIST was calculated from the works of \protect\cite{Moore_1971} \& \protect\cite{Gutmann_1973}. }
  \centering

    \begin{tabular}{cccc} \hline
            \textbf{Level} & \textbf{Config.}      & \TableNewLine{\textbf{$I_p$ NIST} \\ \textbf{/ Ryd}} & \TableNewLine{\textbf{$I_p$ DARC} \\ \textbf{/ Ryd}} \\ \hline
 \noalign{\smallskip}
    1  & 5d$^{10}$6s ($^2$S$_{1/2}$)          & 3.111 & 3.167 \\ \noalign{\smallskip}
    2  & 5d$^{10}$6p ($^2$P$_{1/2}^{\circ}$)  & 2.417 & 2.481 \\ \noalign{\smallskip}
    3  & 5d$^{10}$6p ($^2$P$_{3/2}^{\circ}$)  & 2.225 & 2.173 \\ \noalign{\smallskip}
    4  & 5d$^{9}$6s$^{2}$ ($^2$D$_{5/2}$)     & 2.189 & 2.290 \\ \noalign{\smallskip}
    5  & 5d$^{9}$6s$^{2}$ ($^2$D$_{3/2}$)     & 1.994 & 1.983 \\ \noalign{\smallskip}
    6  & 5d$^{9}$6s6p ($^4$P$_{5/2}^{\circ}$) & 1.595 & 1.474 \\ \noalign{\smallskip}
    7  & 5d$^{9}$6s6p ($^4$F$_{7/2}^{\circ}$) & 1.538 & 1.420 \\ \noalign{\smallskip}
    8  & 5d$^{9}$6s6p ($^4$F$_{5/2}^{\circ}$) & 1.534 & 1.415 \\ \noalign{\smallskip}
    9  & 5d$^{9}$6s6p ($^4$P$_{3/2}^{\circ}$) & 1.513 & 1.389 \\ \noalign{\smallskip}
    10 & 5d$^{10}$6d  ($^2$D$_{3/2}$)         & 1.430 & 1.490 \\ \noalign{\smallskip}
    \hline
    \noalign{\smallskip}
    \end{tabular}%
    \label{tab:PbIV-Ionization_Potential}%
\end{table}

\subsection{Photoionization of \ion{Pb}{v}}

We preserved the 100 lowest lying energy levels from the 2,566 present in our \ion{Pb}{vi} target for the close coupled scattering calculations. The 60 energy levels which had a corresponding experimental value in the literature were calibrated to that value.  The remaining 40 energy levels were shifted either by the mean shift arising from the other levels in the same configuration, or by the mean shift of the 60 calibrated energy levels (1.205\%).  A continuum orbital basis of 25 was selected for each dipole pair included, with a $R$-matrix boundary between the inner and outer region assigned to 13.29au.  The scattering calculations were performed from 0 $\leq J \leq$ 6, resulting in 18 unique dipole pairs and 36 partial waves.  The number of close-coupled channels peaked at 624, with the size of the $N+1$ Hamiltonian matrix reaching 15600 $\times$ 15600.

\begin{figure*}
\centering
\includegraphics[scale=0.35]{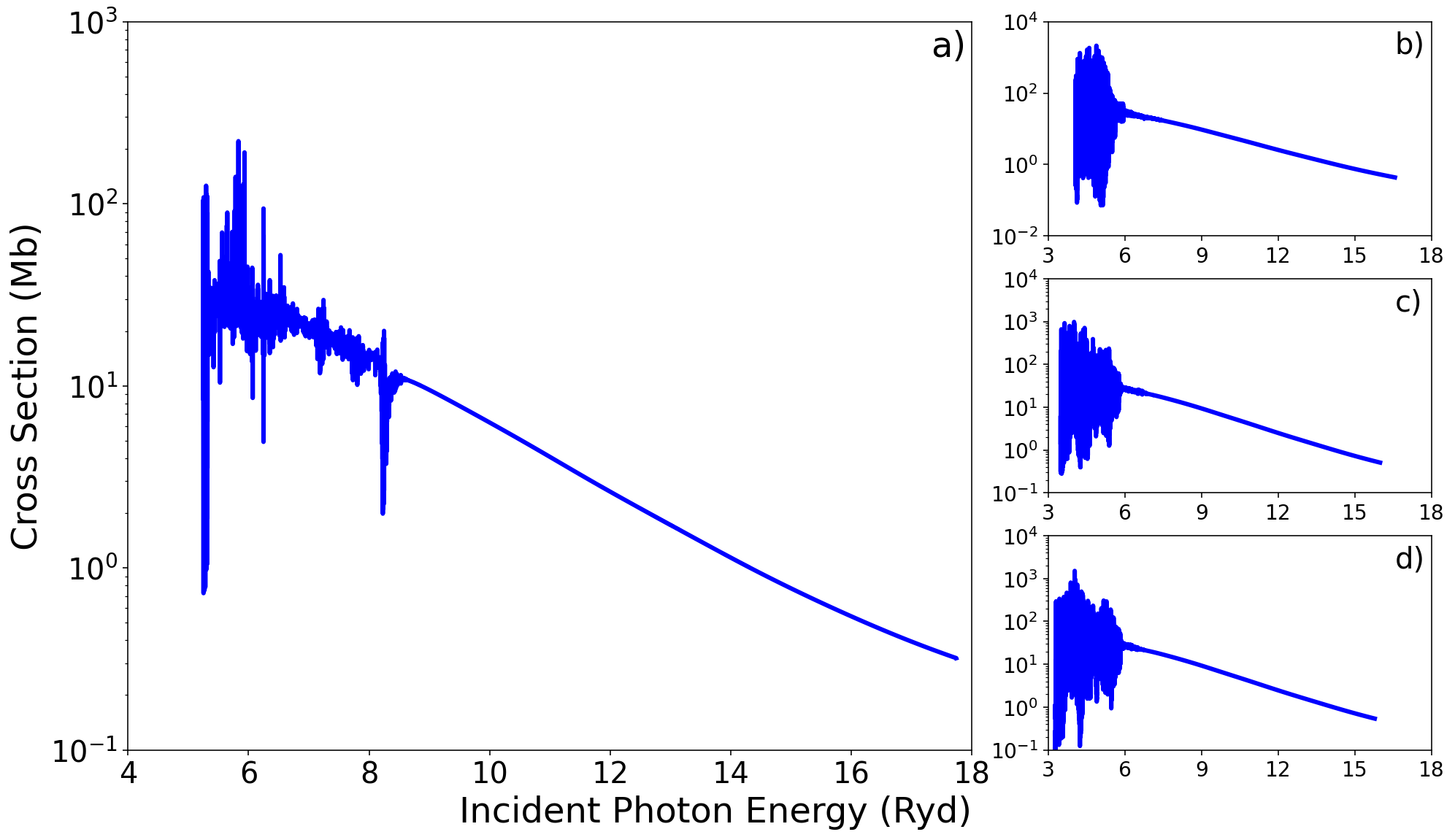}
\caption{Sample photoionization cross sections of \ion{Pb}{v} for Levels a) - 1 (5d$^{10}$ ($^1$S$_0$)), b) - 4 (5$d^{9}$6s ($^3$D$_{1}$)), c) - 7 (5d$^{9}$6p ($^3$F$_{3}^{\circ}$)) and d) - 10 (5d$^{9}$6p ($^3$F$_{4}^{\circ}$)).  Cross sections for the other levels are included in the accompanying {\sc topbase} file.} 
\label{fig:PbV_Cross_Sections}
\end{figure*}

The photon energy mesh grid was defined as having 100,000 points, starting at 0.05 Ryd and increasing in regular intervals of 5$\times$10$^{-6}$ Ryd, which will extend to energies up to 17.8 Ryd.  A sample of photoionization cross sections from both the ground and excited states is illustrated in Fig. \ref{fig:PbV_Cross_Sections}.  The ionization potentials arising from the first 10 levels in \ion{Pb}{v} are displayed in Table \ref{tab:PbV-Ionization_Potential}.  The ionization potentials are in good agreement with those derived from experimental measurements, with the discrepancies between the experimental and computational values being $\leq$ 0.2 Ryd.

\begin{table}
  \caption{The ionization potential arising from our {\sc darc} photoionization calculations for the first 10 levels in \ion{Pb}{v}, compared with their equivalent values in NIST.  The ionization potential of \ion{Pb}{v} is taken to be 5.06 Ryd (\protect\cite{Mack_1935}).  $I_p$ NIST was calculated from the work of \protect\cite{Joshi_1990}.}
  \centering

    \begin{tabular}{cccc} \hline
            \textbf{Level} & \textbf{Config.}      & \TableNewLine{\textbf{$I_p$ NIST} \\ \textbf{/ Ryd}} & \TableNewLine{\textbf{$I_p$ DARC} \\ \textbf{/ Ryd}} \\ \hline
 \noalign{\smallskip}
    1  & 5d$^{10}$  ($^1$S$_0$)          & 5.060 & 5.216  \\ \noalign{\smallskip}
    2  & 5d$^{9}$6s ($^3$D$_3)$          & 4.051 & 4.214  \\ \noalign{\smallskip}
    3  & 5d$^{9}$6s ($^3$D$_2)$          & 4.017 & 4.177  \\ \noalign{\smallskip}
    4  & 5d$^{9}$6s ($^3$D$_1)$          & 3.851 & 4.019  \\ \noalign{\smallskip}
    5  & 5d$^{9}$6s ($^1$D$_2)$          & 3.821 & 3.987  \\ \noalign{\smallskip}
    6  & 5d$^{9}$6p ($^3$P$_2^{\circ}$)  & 3.285 & 3.396  \\ \noalign{\smallskip}
    7  & 5d$^{9}$6p ($^3$F$_3^{\circ}$)  & 3.264 & 3.373  \\ \noalign{\smallskip}
    8  & 5d$^{9}$6p ($^3$F$_2^{\circ}$)  & 3.082 & 3.191  \\ \noalign{\smallskip}
    9  & 5d$^{9}$6p ($^3$P$_1^{\circ}$)  & 3.060 & 3.172  \\ \noalign{\smallskip}
    10 & 5d$^{9}$6p ($^3$F$_4^{\circ}$)  & 3.045 & 3.165  \\ \noalign{\smallskip}
    \hline
    \noalign{\smallskip}
    \end{tabular}%
    \label{tab:PbV-Ionization_Potential}%
\end{table}

\subsection{Photoionization of \ion{Pb}{vi}}

The 1,679 energy levels present in our \ion{Pb}{vii} target were reduced to the lowest lying 157 for the close coupled scattering calculations.  Of these, the 120 levels with corresponding experimental values reported in the literature were shifted, with the remaining 37 levels shifted based on the mean shift arising from other levels from the same configuration.  The $R$-matrix between the inner and outer calculations was set to 13.29au, and a continuum orbital basis of 23 was selected for each dipole pair.  Scattering calculations were performed for 44 partial waves across 22 unique dipole pairs where 0.5 $\leq J \leq$ 7.5.  The number of close coupled channels peaked at 1005, with the size of the $N + 1$ Hamiltonian matrix peaking at 23148 $\times$ 23148.

\begin{figure*}
\centering
\includegraphics[scale=0.35]{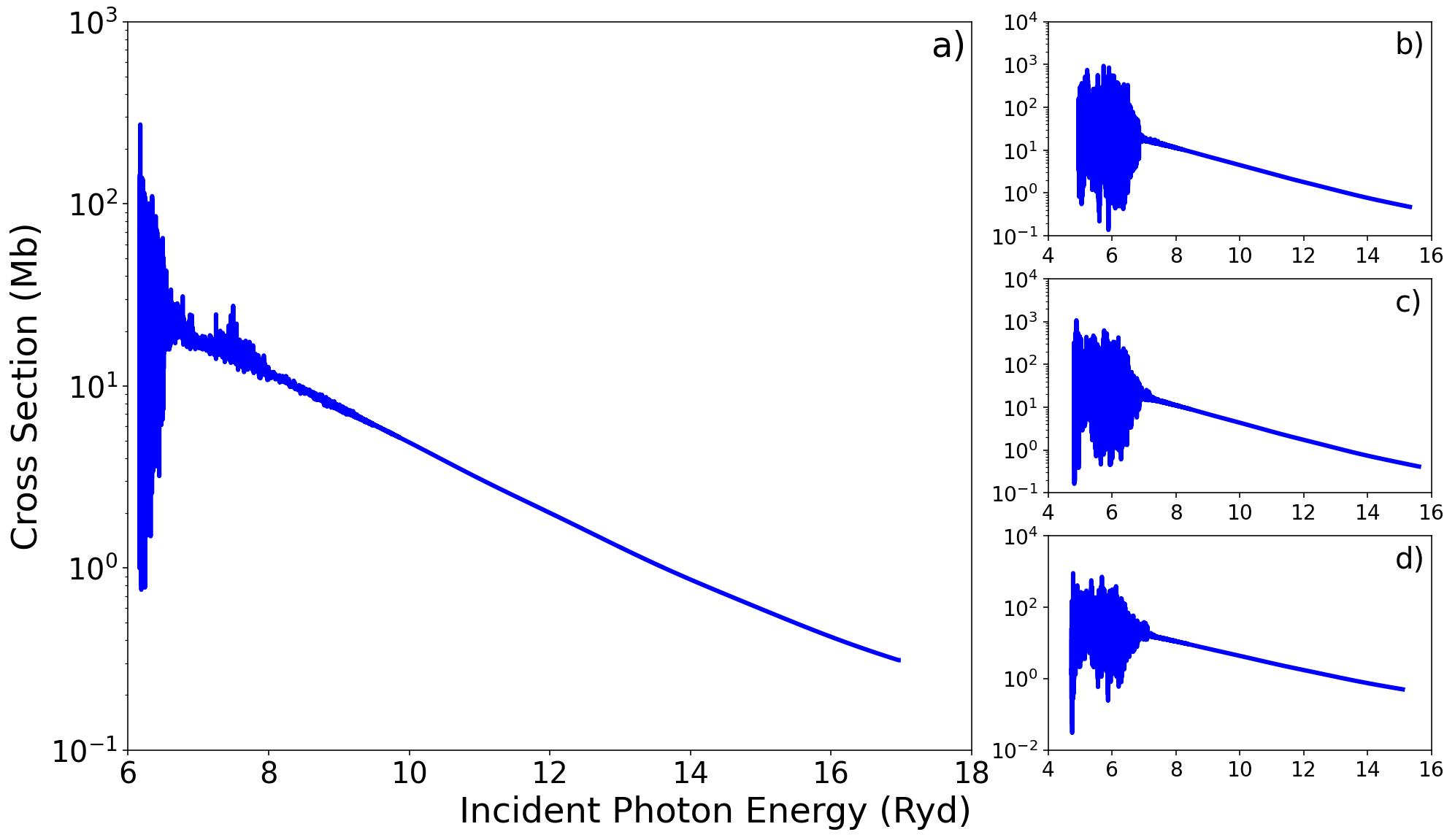}
\caption{Sample photoionization cross sections of \ion{Pb}{vi} for Levels a) - 1 (5d$^9$ ($^2$D$_{5/2}$)) , b) - 4 (5d$^8$6s ($^4$F$_{7/2}$)), c) - 7 (5d$^8$6s ($^4$F$_{5/2}$))  and d) - 10 (5d$^8$6s ($^4$P$_{3/2}$)).  Cross sections for the other levels are included in the accompanying {\sc topbase} file.} 
\label{fig:PbVI_Cross_Sections}
\end{figure*}

The photon energy mesh grid was defined as having 36,000 points, starting at 1$\times$10$^{-5}$ Ryd and increasing in regular intervals of 8$\times$10$^{-6}$ Ryd.  This allowed the photoionization cross sections to be calculated up to energies of $\leq$ 16.9 Ryd.  A sample of the ground and excited photoionization cross sections are shown in Fig. \ref{fig:PbVI_Cross_Sections}.  The ionization potential from the first 10 levels for \ion{Pb}{vi} are displayed in Table \ref{tab:PbVI-Ionization_Potential}.  The ionization potentials are in good agreement with those derived from experimental measurements, with the discrepancies between the experimental and theoretical values being $\leq$ 0.2 Ryd.

\begin{table}
  \caption{The ionization potential arising from our {\sc darc} photoionization calculations for the first 10 levels in \ion{Pb}{vi}, compared with their equivalent values in the literature.  The ionization potential of \ion{Pb}{vi} is taken to be 6.10 Ryd (\protect\cite{Rodrigues_2004}).  $I_p$ Lit. was calculated from the work of \protect\cite{Raassen_1990_PbVI}.}
  \centering

    \begin{tabular}{cccc} \hline
            \textbf{Level} & \textbf{Config.}      & \TableNewLine{\textbf{$I_p$ Lit.} \\ \textbf{/ Ryd}} & \TableNewLine{\textbf{$I_p$ DARC} \\ \textbf{/ Ryd}} \\ \hline
 \noalign{\smallskip}
    1  & 5d$^9$   ($^2$D$_{5/2}$)  & 6.100 & 6.162  \\ \noalign{\smallskip}
    2  & 5d$^9$   ($^2$D$_{3/2}$)  & 5.905 & 5.966  \\ \noalign{\smallskip}
    3  & 5d$^8$6s ($^4$F$_{9/2}$)  & 4.900 & 5.018  \\ \noalign{\smallskip}
    4  & 5d$^8$6s ($^4$F$_{7/2}$)  & 4.836 & 4.951  \\ \noalign{\smallskip}
    5  & 5d$^8$6s ($^2$D$_{5/2}$)  & 4.799 & 4.903  \\ \noalign{\smallskip}
    6  & 5d$^8$6s ($^2$D$_{3/2}$)  & 4.776 & 4.882  \\ \noalign{\smallskip}
    7  & 5d$^8$6s ($^4$F$_{5/2}$)  & 4.681 & 4.804  \\ \noalign{\smallskip}
    8  & 5d$^8$6s ($^2$F$_{7/2}$)  & 4.674 & 4.797  \\ \noalign{\smallskip}
    9  & 5d$^8$6s ($^4$P$_{1/2}$)  & 4.653 & 4.751  \\ \noalign{\smallskip}
    10 & 5d$^8$6s ($^4$P$_{3/2}$)  & 4.614 & 4.729  \\ \noalign{\smallskip}
    \hline
    \noalign{\smallskip}
    \end{tabular}%
    \label{tab:PbVI-Ionization_Potential}%
\end{table}

\begin{figure}
\centering
\includegraphics[width=1.0\columnwidth]{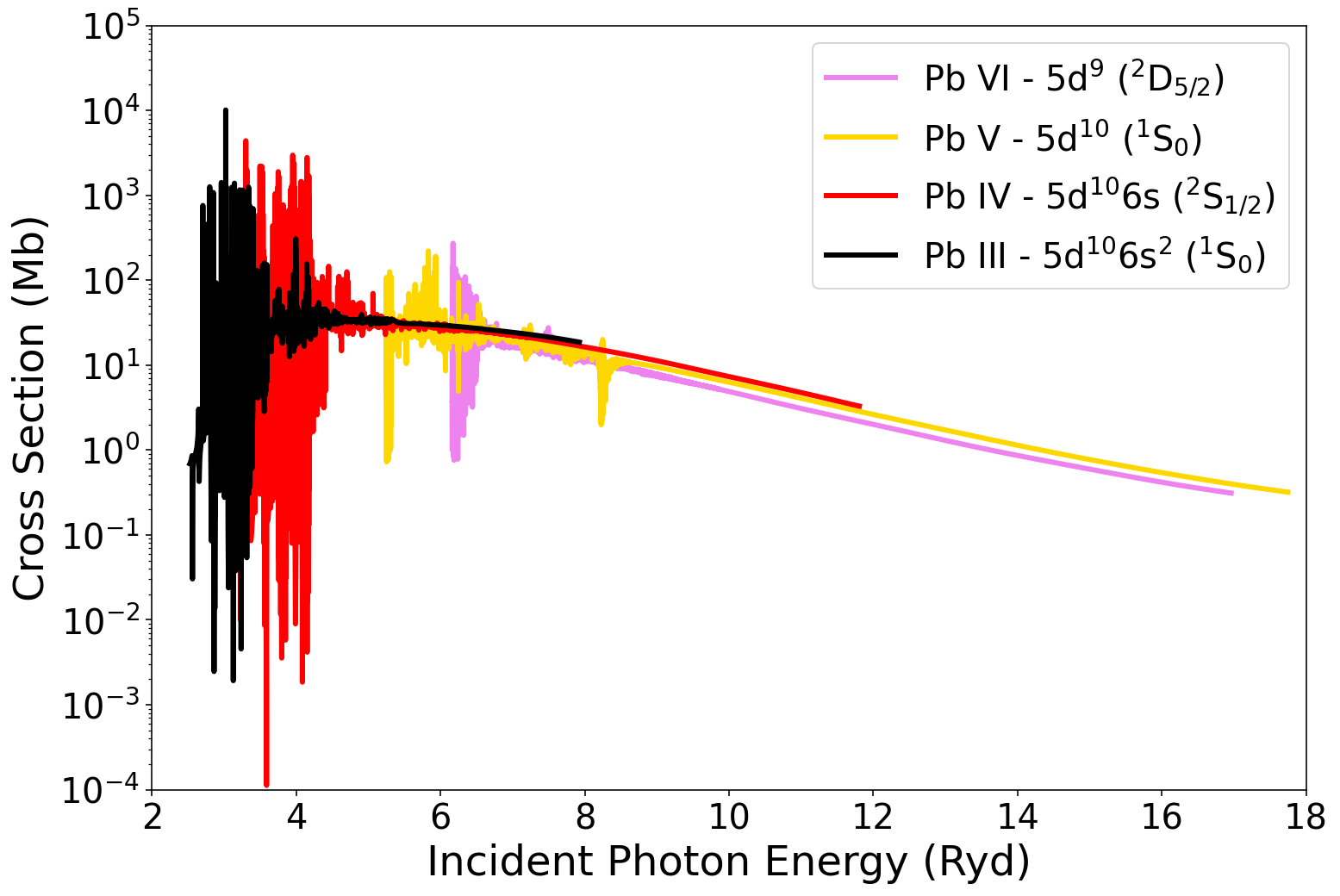}
\caption{The ground state photoionization cross sections across the Pb isonuclear sequence.}
\label{fig:Pb_Isonuc_Comparison}
\end{figure}

We have now discussed the photoionization cross sections for each of the Pb ions.  As a final check of their validity, we present in Fig. \ref{fig:Pb_Isonuc_Comparison} how the magnitudes of our cross sections for the ground state varies along the isonuclear sequence of Pb.  It can be seen that, as the net positive charge of the Pb species increases, the magnitude of the photoionization cross section decreases.  The ionization of successive electrons means that the remaining bound electrons will experience a greater share of electrostatic potential exerted by the nucleus, and hence more energy needs to be applied to remove subsequent electrons.  It therefore seems reasonable that photoionization is occurring at a lower rate for highly charged systems.  This self-consistency among our four data sets provides is a good indication of their reliability.

\section{Application}
\label{sec:Application}

\begin{figure*}
\centering
\includegraphics[width=0.85\textwidth]{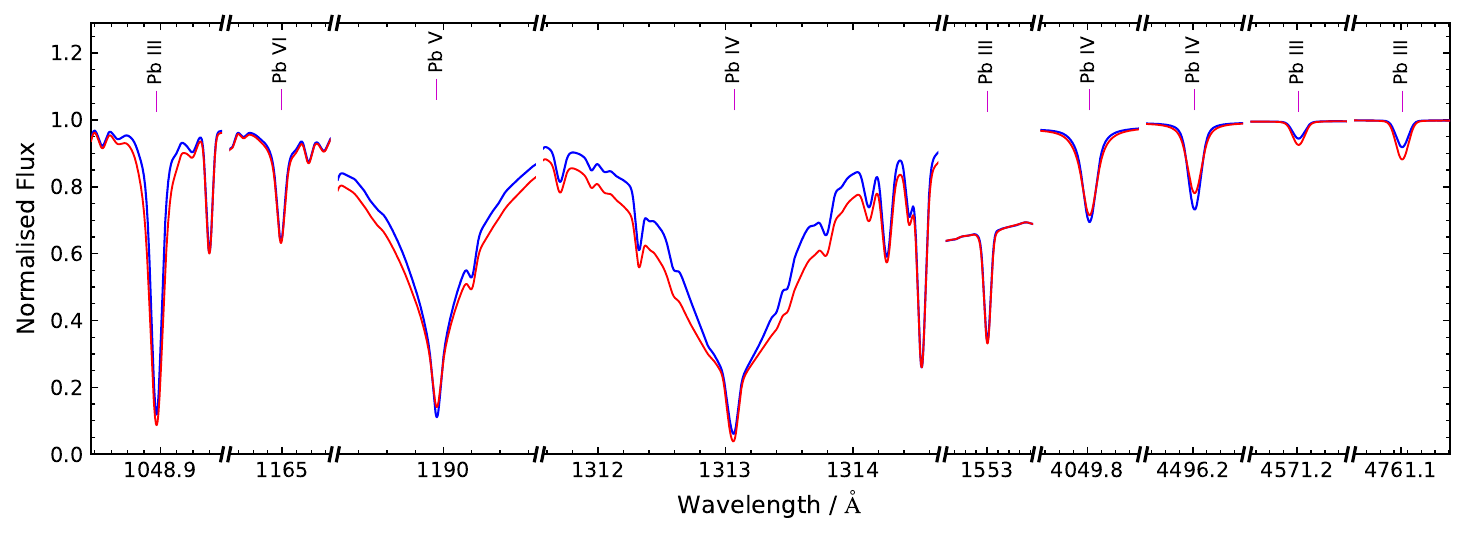}
\vspace{-5pt}
\caption{Predicted lead lines in a \textsc{Tlust/Synspec} model of EC\,22536$-$5304 in LTE (red) and non-LTE (blue), using the atomic data computed here. Both models were convolved to a resolving power of $R = 40\,000$ for clarity. 
}
\label{fig:Pb_nLTE_UV}
\end{figure*}



Accurate photoionization data for multiply ionized lead are important for lead abundance measurements in O- and B-type stars, since such models must be computed without assuming LTE. The \ion{Pb}{iv} resonance lines at 1028 and 1313\,\AA\ are detectable in B-type stars even at solar abundances, while in lead-rich stars, such as the hot subdwarf EC\,22536$-$5304 \citep{Dorsch_2021}, numerous additional \ion{Pb}{iii-vi} lines are observed in the optical and ultra-violet (UV). We assessed the impact of the new atomic data by computing non-LTE level populations of \ion{Pb}{iii-vi} in a \textsc{Tlusty} model atmosphere of EC\,22536$-$5304 ($T_\mathrm{eff}=37\,750$\,K, $\log g = 5.8$), and comparing with a model where lead was included only in LTE spectrum synthesis. 
As shown in Fig.~\ref{fig:Pb_nLTE_UV}, non-LTE effects significantly alter both UV and optical lead lines, leading to weaker and sharper profiles compared to LTE. 
These effects arise from substantial deviations from LTE in both ionisation balance (Fig.~\ref{fig:Pb_ion}, left) and detailed level populations (Fig.~\ref{fig:Pb_departure}). 
\ion{Pb}{iii-iv} are depleted and \ion{Pb}{v-vi} slightly enhanced in the outer photosphere, with ground states of \ion{Pb}{iii-v} more strongly populated relative to excited states.
Because the atmosphere of EC\,22536$-$5304 is extremely enriched in lead \citep[$\sim$0.01\,\% by number, or $10^{6.3}$ times solar;][]{Dorsch_2021}, the additional lead opacity also affects its atmospheric structure, cooling the line-forming region by a maximum of about 175\,K (Fig.\ \ref{fig:Pb_ion}, right).
A non-LTE treatment of lead is therefore essential for reliable abundance determinations.

Abundance measurements are not the only application for our data.  One hypothesis explaining the high Pb absorption in the spectra of heavy metal subdwarfs, such as EC\,22536$-$5304, is abundance stratification \citep[cf.][]{Scott_2024}. This may be caused by the competing processes of gravitational settling and radiative levitation concentrating Pb into the line-forming region of the atmosphere. To calculate radiative levitation forces, the opacities of the diffusing ions in the atmosphere must be included, as in Eq.~3 of \citet{Schuh2002}. Whilst previous work \citep{Alonso-Medina_2011,Colon_2014,Safronova_2004} allowed the calculation of Pb line opacities, the full opacity requires the inclusion of bound-free transitions, which is now possible thanks to the new photoionization cross sections presented in this work.

\section{Conclusions}
\label{sec:Conclusions}

Many AGB and sdO/B stars have been observed to be enriched with Pb, with corresponding absorption lines originating from a range of Pb charge states.  We have aimed to provide accurate and extensive level-resolved photoionization cross sections of commonly observed Pb species in hot stars, and updated the available energy levels, Einstein A-coefficients and oscillator strengths for these species.  This is to assist in the simulation and modelling of the stellar spectra observed from these stars, which may provide a more comprehensive understanding of their evolutionary pathways, and the synthesis of heavy metals. 

New atomic structure models for \ion{Pb}{iv}, {\sc v}, {\sc vi} and {\sc vii} have been developed using the {\sc grasp}$^0$ {\sc fortran} package.  The energy levels, Einstein A-coefficients and oscillator strengths were found to be in very good agreement with experimental work presented in NIST, and with other previously published work.  Photoionization cross sections for \ion{Pb}{iii}, {\sc iv}, {\sc v} and {\sc vi} were calculated using the {\sc darc} suite of codes.  We present level-resolved ground and excited state cross sections for the four Pb species, and while there is no equivalent experimental or theoretical work with which to compare, the magnitudes of the individual cross sections are in alignment with what would be expected for an isonuclear sequence.  The finalized cross sections are presented in {\sc topbase} format.

The Pb data set allows for the modelling of Pb abundances within O- and B- type stars under non-LTE conditions.  To explore the extent to which this may affect current LTE stellar models, we examined a test case using the Pb enriched subdwarf star EC\,22536$-$5304.  It was seen that there are noticeable differences in the Pb line profiles under LTE and non-LTE conditions, arsing from variations in the electron level populations and lead ionization fraction models.  Different Pb species were depleted or enriched in the line forming regions in the transition to non-LTE conditions.  This confirms the importance of incorporating non-LTE conditions into the stellar modelling for accurate stellar abundance calculations.

\section*{Acknowledgements}

We thank Prof.\ Simon Jeffrey of the Armagh Observatory and Planetarium for helpful discussions.  We are grateful for the use of the computing resources from the Northern Ireland High Performance Computing (NI-HPC) service funded by EPSRC (EP/T022175).  DJD thanks the Science and Technology Facilities Council (STFC) of the UK Research and Innovation (UKRI) body for their support through his studentship.
MD was supported by the Deutsches Zentrum für Luft- und Raumfahrt (DLR) through grant 50-OR-2304. 

\section*{Data Availability}

The \ion{Pb}{iii}, {\sc iv}, {\sc v} and {\sc vi} photoionization cross sections described in this work are available in {\sc topbase} format and can be obtained at \cite{Ballance_2025}.  Other data relating to this work can be provided by the corresponding author upon reasonable request.






\begin{appendix}

\section{Additional Figures}

\begin{figure*}
\centering
\includegraphics[width=0.49\textwidth]{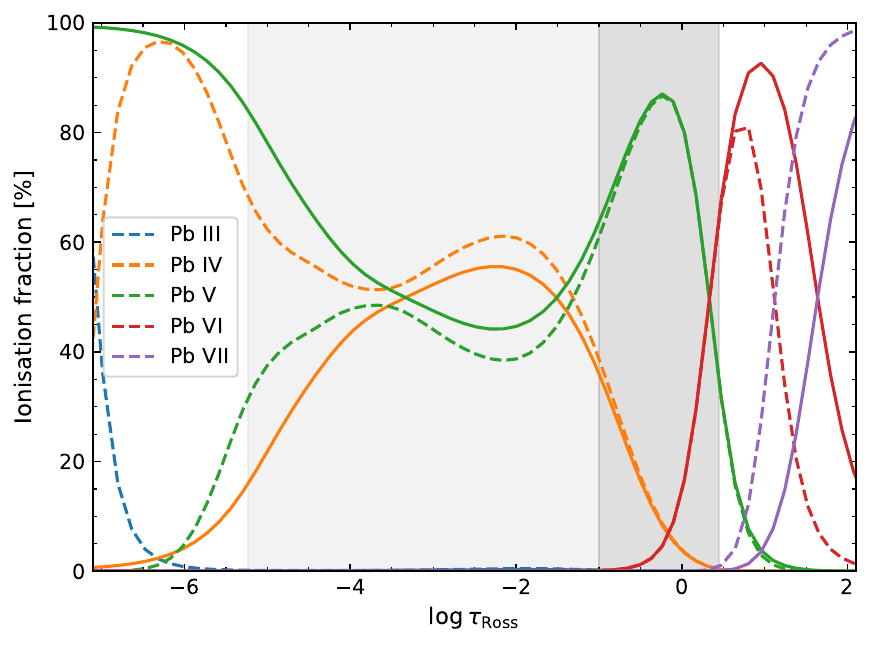}
\includegraphics[width=0.49\textwidth]{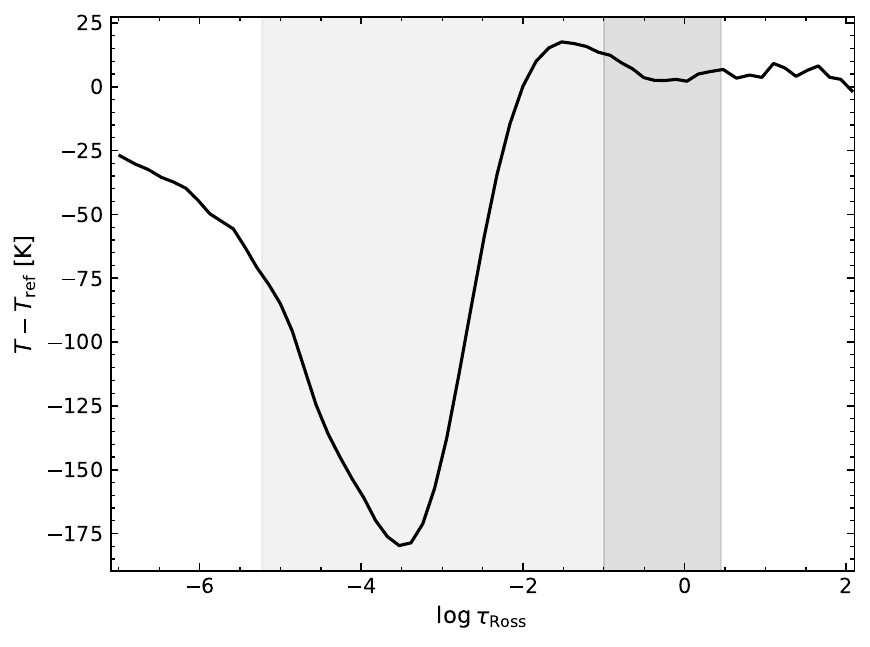}
\caption{\textit{Left}: Lead ionization fractions in models of EC\,22536$-$5304 as a function of Rosseland mean optical depth $\tau_\mathrm{Ross}$, comparing LTE (dashed) and non-LTE (solid) with the atomic data computed here. 
In non-LTE, \ion{Pb}{iii-iv} is slightly suppressed, while \ion{Pb}{v-vi} is enhanced. 
\textit{Right}: Temperature difference between models of EC\,22536$-$5304 with ($T$) and without Pb ($T_\mathrm{ref}$). Pb opacity slighly cools the line-forming region (light grey), while the continuum-forming region (dark grey) remains unaffected. All models also include opacity from H, He, C, N, O, Ne, Mg, Al, Si, S, Fe, and Ni at abundances appropriate for the star.
}
\label{fig:Pb_ion}
\end{figure*}

\begin{figure*}
\centering
\includegraphics[width=0.49\textwidth]{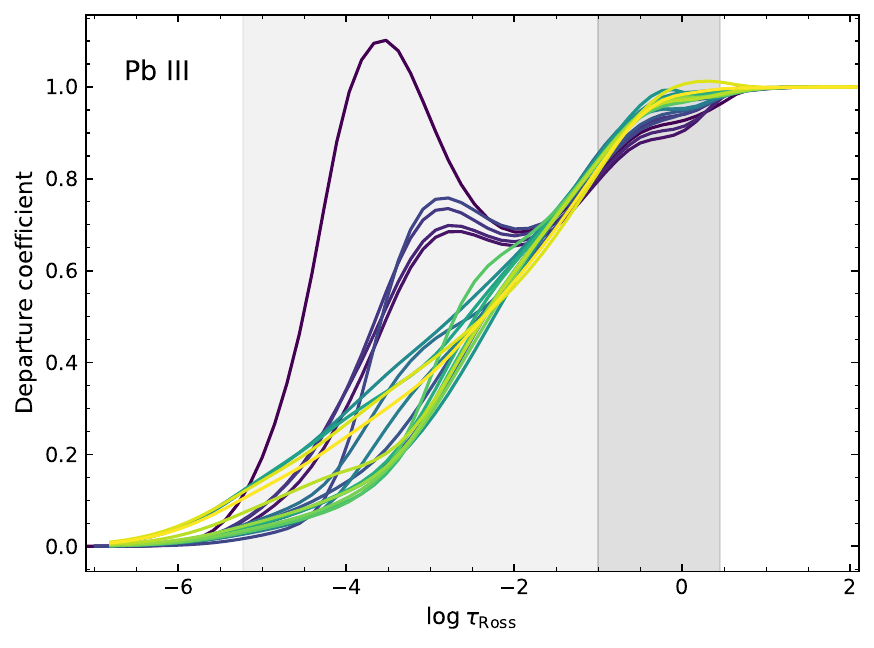}
\includegraphics[width=0.49\textwidth]{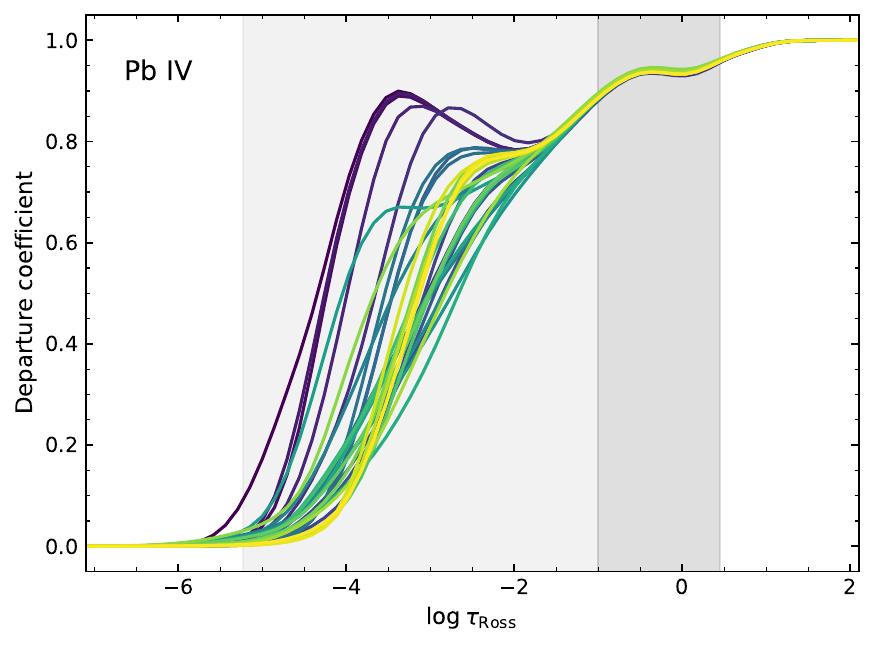}\\
\includegraphics[width=0.49\textwidth]{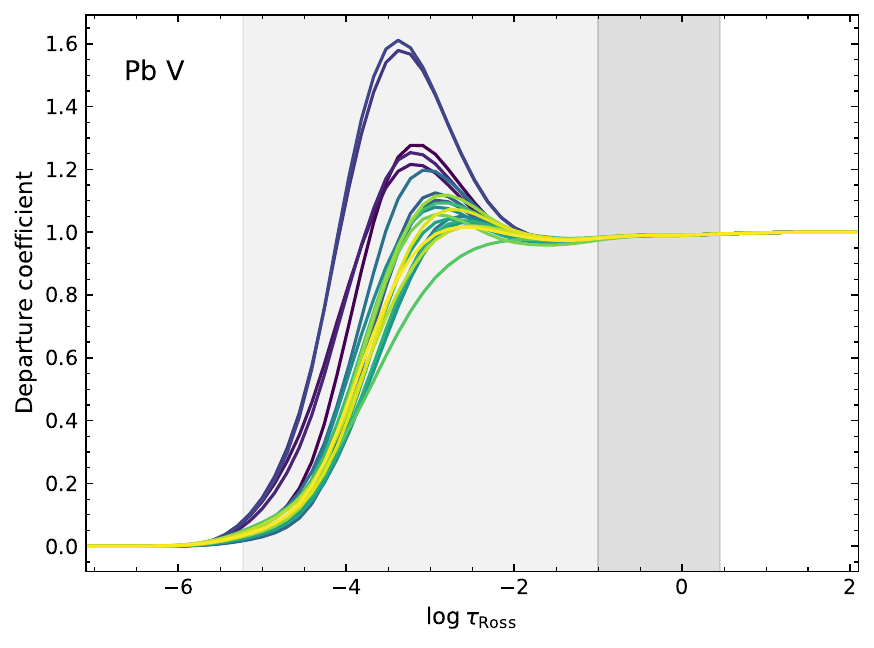}
\includegraphics[width=0.49\textwidth]{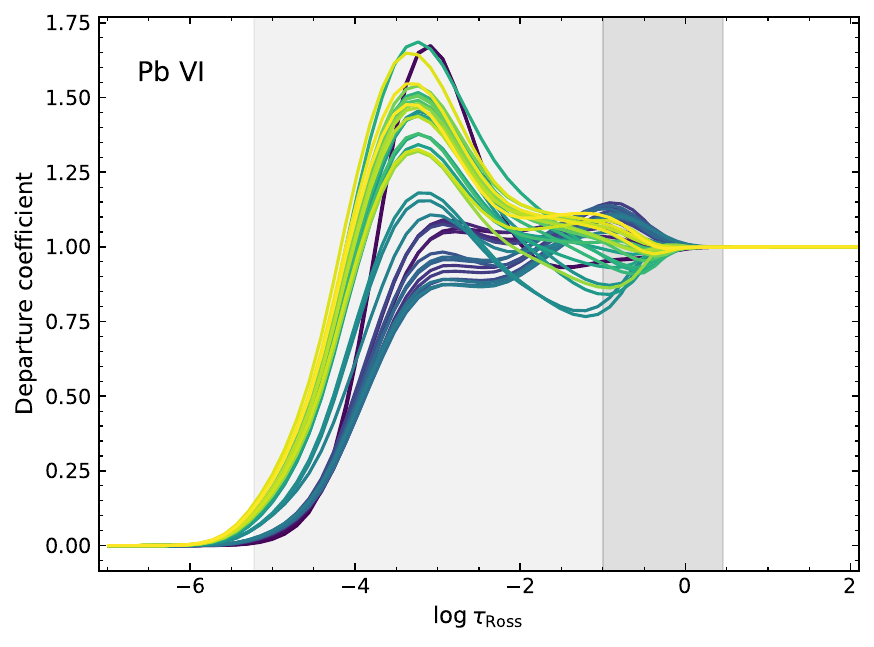}
\vspace{-5pt}
\caption{Departures from LTE for energy level populations in a model of EC\,22536$-$5304, using the data provided here. The light-shaded region indicates where spectral lines form, and the dark-shaded region marks the continuum-forming layer. The lowest-lying levels are purple, the highest yellow.
}
\label{fig:Pb_departure}
\end{figure*}




\end{appendix}


\bsp	
\label{lastpage}
\end{document}